\documentclass[12pt]{article}
\pdfoutput=1

\usepackage{draft}
\usepackage{putex} 
\usepackage[all,cmtip]{xy}
\usepackage{fancybox}
\usepackage{subcaption}
\usepackage{graphicx}
\usepackage{cite}
\usepackage{mciteplus}
\usepackage{skak}
\usepackage{amsmath}
\DeclareFontFamily{OT1}{pzc}{}
\DeclareFontShape{OT1}{pzc}{m}{it}{<-> s * [1.10] pzcmi7t}{}
\DeclareMathAlphabet{\mathpzc}{OT1}{pzc}{m}{it}

\def\({\left(}
\def\){\right)}

\usepackage{tikz-cd}

\usepackage{latexsym}
\usepackage{stmaryrd}

\usepackage{cite}
\usepackage{framed}
\definecolor{shadecolor}{rgb}{0.95,0.95,0.97}
\definecolor{refkey}{rgb}{0.5,0.5,0}
\definecolor{labelkey}{rgb}{0.5,0.5,0}
\definecolor{citekey}{rgb}{0.5,0.5,0}
\definecolor{darkgreen}{rgb}{0,0.5,0}
\definecolor{darkblue}{cmyk}{0.9,0.9,0,0}
\definecolor{darkred}{rgb}{0.6,0,0.3}
\usepackage{ifpdf}
\ifpdf
\usepackage{graphicx} 
\usepackage{epsfig}
\usepackage[setpagesize=false,pagebackref=false, linktocpage, bookmarksopen=true, colorlinks=true, linkcolor=blue,citecolor=blue,urlcolor=blue]{hyperref}
\usepackage{hyperref}

\else
\usepackage{graphicx}

\fi

\newcommand{\ep}{\epsilon}
\newcommand{\ve}{\varepsilon}  
\newcommand{\bD}{\bm \Delta}

\def\del{\partial}

\def\fn#1{\footnote{#1}}
\def\nn{\nonumber}
\def\eqref#1{(\ref{#1})}
\def\comma{\,,}
\def\period{\,.}

\def\bm#1{\text{\boldmath $#1$}}
\def\ket#1{|#1\rangle}
\def\bra#1{\langle #1 |}

\def\Xint#1{\mathchoice
	{\XXint\displaystyle\textstyle{#1}}%
	{\XXint\textstyle\scriptstyle{#1}}%
	{\XXint\scriptstyle\scriptscriptstyle{#1}}%
	{\XXint\scriptscriptstyle\scriptscriptstyle{#1}}%
	\!\int}
\def\XXint#1#2#3{{\setbox0=\hbox{$#1{#2#3}{\int}$}
		\vcenter{\hbox{$#2#3$}}\kern-.5\wd0}}

\def\dashint{\Xint-}
\newcommand{\beq}{\begin{equation}}
\newcommand{\eeq}{\end{equation}}
\def\red#1{\textcolor[rgb]{1, 0, 0}{#1}}

\def\nullify#1{}

\makeatletter
\def\section{\@startsection {section}{1}{\z@}{-3.5ex plus -1ex minus 
		-.2ex}{2.3ex plus .2ex}{\large\bf}}
\makeatother
\makeatletter
\def\subsection{\@startsection {subsection}{1}{\z@}{-3.5ex plus -1ex minus 
		-.2ex}{2.3ex plus .2ex}{\normalsize\bf}}
\makeatother

\begin{document}
	
	\preprint{}

	\institution{IAS}{School of Natural Sciences, Institute for Advanced Study, 
		Princeton, NJ 08540, USA}
	\institution{PU}{Joseph Henry Laboratories, Princeton University, Princeton, NJ 08544, USA}
	\institution{CMSA}{Center of Mathematical Sciences and Applications, Harvard University, Cambridge, MA 02138, USA}
	\institution{HU}{Jefferson Physical Laboratory, Harvard University,
		Cambridge, MA 02138, USA}

	\title{
	Non-perturbative Defect One-Point Functions \\in Planar $\cN=4$ Super-Yang-Mills
	}

	\authors{Shota Komatsu\worksat{\IAS}    and Yifan Wang\worksat{\PU,\CMSA,\HU}}

	\abstract{
			The  four dimensional $\cN=4$ super-Yang-Mills (SYM) theory exhibits rich dynamics in the presence of codimension-one conformal defects.  The new structure constants of the extended operator algebra consist of one-point functions of local operators which are nonvanishing due to the defect insertion and carry nontrivial coupling dependence. We study an important class of half-BPS superconformal defects engineered by D5 branes that share three common directions with the D3 branes and involve Nahm pole configurations for the SYM fields on the D3 brane worldvolume. 
			In the planar large $N$ limit, we obtain non-perturbative results in the 't Hooft coupling $\lambda$ for the defect one-point functions of both BPS and non-BPS operators, building upon recent progress in localization \cite{Wang:2020seq} and integrability methods \cite{Jiang:2019xdz,Jiang:2019zig}.
			
			For BPS operator insertions in the SYM with D5-brane type boundary or interface, we derive an effective two dimensional defect-Yang-Mills (dYM) theory from supersymmetric localization, 
			which gives an efficient way to extract defect observables and generates a novel matrix model for the defect one-point function. By solving the matrix model in the large $N$ limit, we obtain exact results in $\lambda$ which interpolate between perturbative Feynman diagram contributions in the weak coupling limit and IIB string theory predictions on $AdS_5\times S^5$ in the strong coupling regime,  providing a precision test of AdS/CFT with interface defects.  
			For general non-BPS operators, we
			develop a non-perturbative bootstrap-type program for integrable boundary states on the worldsheet of the IIB string theory, corresponding to the interface defects in the planar SYM. Such integrable boundary states are constrained by a set of general consistency conditions for which we present explicit solutions that 
			reproduce and extend the known results at weak coupling from integrable spin-chain methods.		
	}
	\date{}

	\maketitle
	
	\tableofcontents
	
	\section{Introduction}
	
	Quantum field theories are known to exhibit rich dynamics in the presence of co-dimension one defects such as boundaries or domain walls (interfaces). In particular topological field theories which do not carry any propagating degrees of freedom in the bulk at all, may harbor nontrivial interactions on the boundary or domain wall. A canonical example of this involves the three-dimensional Chern-Simons theory on a manifold with boundary and the corresponding (chiral) Wess-Zumino-Witten model (or general rational conformal field theories) on the boundary \cite{Witten:1988hf,Moore:1989yh,Elitzur:1989nr}, as a special instance of the anomaly inflow mechanism \cite{Callan:1984sa}. More interestingly is when the bulk theory is also strongly interacting. In such cases, the anomaly arguments alone are not enough to pin down the dynamics of the coupled system. Fortunately when the bulk theory enjoys conformal symmetry, as is often the case for strongly-coupled fixed points of renormalization-group (RG) flows, there are direct methods to determine the physical observables non-perturbatively, which broadly speaking belong to the \textit{conformal bootstrap program} \cite{Poland:2018epd}. This is thanks to an axiomatic definition of the fixed point theory, a conformal field theory (CFT), in terms of the spectrum of local operators and the operator-product-expansion (OPE) of correlation functions. The conformal symmetry, crossing symmetry together with unitarity imply an infinite set of constraints for the spectrum and OPE coefficients (three-point-functions), which can be explored systematically by bootstrap methods. The bootstrap problem for local operators and their correlation functions has a natural extension in the presence of boundaries or domain walls preserving the the conformal subalgebra longitudinal to the defect \cite{Liendo:2012hy,Liendo:2016ymz,Billo:2016cpy}. Compared to the case without the defects, we now have a richer setup with additional structure constants intrinsic to the defects, known as the defect one-point functions of bulk local operators.  Together with the OPE expansion of the local operators, they determine completely the local correlation functions in the presence of the conformal defect. Furthermore, these defect one-point functions constrain the spectrum of local operators that are confined to the world-volume of the conformal defect. This is achieved for example by studying a different bulk-boundary OPE limit of the two-point function of bulk local operators that exchanges boundary operators in the intermediate channel and deriving a crossing equation that relates to the OPE limit of the local operators in the bulk.

	In four spacetime dimensions, a large class of  strongly coupled fixed points are produced by RG flows from Yang-Mills theories coupled to matter fields in various representations, which admit interesting boundary and domain wall dynamics. Among such four-dimensional CFTs, the 
	$\cN=4$ super Yang-Mills (SYM) holds a special place. On one hand, it shares many features of the other strongly-coupled gauge theories. On the other hand, the fact that it has the maximal supersymmetry allows for a number of analytic methods to probe its dynamics, such as supersymmetric localization  \cite{Nekrasov:2002qd,Pestun:2007rz,Pestun:2016zxk}, integrability \cite{Beisert:2010jr}, and superconformal bootstrap (a refinement of the ordinary bootstrap program outlined above) \cite{Poland:2018epd}. Furthermore, via the conjectured holographic correspondence, observables in the  $\cN=4$ SYM in the large $N$ limit are mapped to those in the type IIB string theory on $AdS_5\times S^5$ and vice versa \cite{Maldacena:1997re,Witten:1998qj,Gubser:1998bc}, thus providing a channel to probe and test quantum gravity using field theory methods. 
	
	The $\cN=4$ SYM is known to host a large family of boundary and interface defects, and the half-BPS ones have been classified in \cite{Gaiotto:2008sa,Gaiotto:2008sd,Gaiotto:2008ak}. They correspond to D5- and NS5-branes (more generally $(p,q)$ five-branes) that share three common directions with the stack of D3-branes that engineer the SYM. The codimension one defect often contains strongly-coupled three-dimensional excitations on its worldvolume, which may be described, via the mirror symmetry (or mirror duality), by RG flows from a three-dimensional $\cN=4$ supersymmetric quiver gauge theory coupled to the SYM \cite{Gaiotto:2008ak}.\footnote{See \cite{DiPietro:2019hqe} (and also \cite{Herzog:2017xha}) for a recent study of interacting boundary CFTs in a non-supersymmetric setting, where the bulk is described by the 4d Maxwell theory.}  
	
	In this paper, we study the  $\cN=4$ super Yang-Mills  in the presence of half-BPS boundary and interface defects of the D5-brane type, building upon recent progress in supersymmetric localization \cite{Wang:2020seq} and integrability methods \cite{Jiang:2019xdz,Jiang:2019zig} to extract  the basic structure constants, the defect one-point function $\la \cO\ra_\cD$ of 
	single-trace local operators in the SYM. 
	Combined with the superconformal bootstrap method \cite{Liendo:2012hy,Liendo:2016ymz}, our results provide  a way to potentially solve the $\cN=4$ SYM in the presence of these interface defects. 
	
	From the general discussion in \cite{Wang:2020seq}, the $\cN=4$ SYM in the presence of half-BPS boundary or interface defects contains a solvable 2d/1d subsector described by (constrained) two-dimensional Yang-Mills theory \cite{Pestun:2009nn} coupled to certain one-dimensional topological quantum mechanics \cite{Dedushenko:2016jxl}, known as the defect-Yang-Mills (dYM). A large class of defect observables in the SYM that preserve a common supercharge $\cQ$  have simple descriptions in the dYM sector and their correlation functions can be extracted using standard two-dimensional gauge theory methods \cite{Witten:1991we,Rusakov:1990rs,Blau:1991mp,Witten:1992xu}.\footnote{See also \cite{Blau:1993hj,Cordes:1994fc} for reviews on this subject.} For the D5-brane interface that interpolate between $U(N)$ and $U(N+k)$ SYM theories for $k\geq 0$, we determine explicitly the dYM sector in this paper. While the D5-brane interface for $k=0$ has a simple Lagrangian description, as a transparent interface stacked with a bifundamental hypermultiplet on its worldvolume, this is not the case when $k>0$ which involve the singular Nahm pole boundary condition \cite{Gaiotto:2008sa}. Thanks to the S-duality of the bulk SYM theory as well as the related mirror symmetry acting on the boundary conditions (boundary theories), we are able to derive the one-dimensional topological quantum mechanics corresponding to the D5-brane interface (boundary) in the dYM. Using two-dimensional gauge theory techniques in the dYM effective theory, the computation of the defect one-point function  $\la \cO \ra_\cD$ is reduced to a single-matrix integral. Compared to the simple Gaussian matrix model familiar for SYM, our matrix model involves a novel single-eigenvalue potential, which comes from the D5-brane defect. 
	By solving this matrix model in the planar large $N$ limit, we determine the one-point functions $\la \cO \ra_\cD$ as exact functions of the 't Hooft coupling $\lambda=N g_4^2$. Expanding the answer in both weak and strong coupling regimes,  our exact result  bridges perturbative answers from Feynman diagram computations in the SYM and holographic results from type IIB string theory on $AdS_5\times S^5$ where the interface is described by a probe D5-brane along an $AdS_4\times S^2$ submanifold with $k$-units of worldvolume flux threading the $S^2$ factor \cite{Karch:2000gx}.

The planar $\cN=4$ SYM is integrable which allows for determination of defect one-point functions using spin-chain methods \cite{Beisert:2010jr,deLeeuw:2015hxa}. Local operators and defect observables are represented by quantum states of the spin chain. In particular the interface defect corresponds to a matrix product state (MPS)   and the bulk local operators correspond to~Bethe eigenstates of the spin chain Hamiltonian. The defect one-point functions $\la\cO\ra_\cD$ is given by the overlap between the MPS and Bethe states \cite{deLeeuw:2015hxa,Buhl-Mortensen:2015gfd,Buhl-Mortensen:2016pxs,deLeeuw:2016umh,Buhl-Mortensen:2016jqo,deLeeuw:2016ofj,Buhl-Mortensen:2017ind,deLeeuw:2017cop,deLeeuw:2018mkd,Grau:2018keb,Gimenez-Grau:2019fld,deLeeuw:2019ebw,Widen:2018nnu}. However, this computation relies on a weak coupling  expansion and is not suitable for accessing correlators in the $\cN=4$ SYM at finite $\lambda$. 
In this paper, we develop a bootstrap-type program for the defect one-point functions at finite $\lambda$ following the general strategy laid out in \cite{Jiang:2019xdz,Jiang:2019zig}. We identify the D5-brane interface defect with a integrable boundary state on the worldsheet of the IIB string theory in the holographic dual, and compute it non-perturbatively by imposing a set of consistency conditions: the Watson's equation, boundary Yang-Baxter equations, and the crossing equation. Knowledge of the integrable boundary states has powerful consequences: it determines the defect structure constants $\la\cO\ra_\cD$ for $\cO$ that are non-BPS. We provide explicit solutions to the consistency conditions and check that they reproduce the known results at weak coupling.
	
	The rest of the paper is organized as follows. We start by reviewing half-BPS interface defects in the $\cN=4$ SYM in Section~\ref{sec:D5mm}, and determine the dYM sector as well as the corresponding matrix model for D5-brane  interface defects using S-duality and mirror symmetry. In Section~\ref{sec:D5onepf}, we determine the defect structure constants $\la \cO\ra_\cD$ of half-BPS operators using the interface matrix model in the large $N$ limit and compare to results from perturbation theory and holographic computations. In Section~\ref{sec:integrable}, we present the bootstrap approach to the defect one-point function of non-BPS operators using integrability. We end by a brief summary and discussion in Section~\ref{sec:discussion}.   
	
	\section{The Matrix Model for D5-brane Interface}
	\label{sec:D5mm}
	
	\subsection{Review of boundary and interface defects in SYM}
	\label{sec:gendef}
	The   4d $\cN=4$ SYM with gauge group $G$ (its Lie algebra denoted by $\mf{g}$) is realized by the dimensional reduction of the 10d SYM. It is thus natural to write its action in terms of the 10d fields as
	\ie
	S_{\rm SYM} = -{1\over 2 g_4^2}\int_{\mR^4} d^4 x\,  \tr \Bigg(
	{1\over 2}F_{MN}F^{MN}-\Psi \Gamma^M D_M\Psi  
	\Bigg)\,.
	\label{SYMact}
	\fe
	Here $M=1,2,\dots,8,9,0$ are 10d spacetime indices which splits into 4d spacetime and R-symmetry (internal) indices $(\m,I)$ with $\m=1,2,3,4$ and $I=5,\dots,9,0$. The bosonic field $A_M$ contains the 4d gauge field $A_\m$ and scalars $\Phi_I$.
	The gaugino $\Psi$ transforms as a chiral spinor of $Spin(10)$ and $\Gamma_M=\{\Gamma_\m,\Gamma_I\}$ are 10d chiral Gamma matrices. 
	We follow the convention of \cite{Pestun:2009nn} for the covariant derivative $D\equiv d+A$ and curvature $F=dA+A\wedge A$. In terms of its $\mf{g}$ components, $A_M\equiv A_M^a T_a$ comes with real coefficients $A_M^a$ and \textit{anti-hermitian} generators $T_a$ of  $\mf{g}$. The trace $\tr(\cdot,\cdot)$ is the Killing form of $\mf{g}$ and is related to the usual trace in a particular representation $R$ by $\tr={1\over 2T_R}\tr_R$, where $T_R$ denotes the Dynkin index of $R$. For $\mf{g}=\mf{su}(N)$, this is identical to the trace in the fundamental representation $\tr=\tr_F$. Finally the generators $T^a$ are normalized by $\tr(T_a T_b)=-{1\over 2}\D_{ab}$. We set the four-dimensional theta angle $\theta=0$ in this paper.

	The superconformal symmetry of the SYM comes from the conformal Killing spinor
	\ie
	\ve = \ep_s + x^\m \Gamma_\m \ep_c
	\fe
	where $\ep_s$ and $\ep_c$ are constant 16-component spinors that correspond to the Poincar\'e and conformal supercharges in the superconformal algebra $\mf{psu}(2,2|4)$. The superconformal transformation of the SYM fields are generated by
	\ie
	&\D_\ve A_M= \varepsilon \Gamma_M \Psi  \,,
	\\
	&\D_\ve\Psi ={1\over 2}F_{MN}\Gamma^{MN}\varepsilon+{1\over 2}\Gamma_{\m I}\Phi^I \partial^\m \varepsilon  \,.
	\label{SUSYos}
	\fe
	
	The half-BPS interface or boundary defect along the hyperplane $x_1=0$ preserve a superconformal subalgebra $\mf{osp}(4|4,\mR)\subset \mf{psu}(2,2|4)$. There is a family of such subalgebras related by inner automorphisms (conjugation) as well as outer-automorphisms ($U(1)_Y$) of $\mf{psu}(2,2|4)$. Here we follow \cite{Wang:2020seq} and fixes this ambiguity by specifying the supercharges preserved by the   defect
	\ie
	\Gamma_{1890} \ep_s=\ep_s,\quad \Gamma_{1890} \ep_c=\ep_c\,.
	\fe
	The corresponding half-BPS subalgebra contains the following bosonic algebras
	\ie
	\mf{osp}(4|4,\mR) \supset \mf{so}(3)_{567} \oplus  \mf{so}(3)_{890} \oplus  \mf{so}(3,2)_{\rm conf}\,.
	\fe
	Here $\mf{so}(3,2)_{\rm conf}$ is the conformal symmetry along the defect, 
	and $\mf{so}(3)_{567} \oplus  \mf{so}(3)_{890}$   a maximal subalgebra of the full $\mf{so}(6)_R$ symmetry.
	
	The 4d $\cN=4$ vector multiplet naturally decomposes with respect to this  $\mf{osp}(4|4,\mR)$ subalgebra into the following 3d $\cN=4$ multiplets
	\ie
	&{\rm hypermultiplet}~:\Psi_-, A_1, X_i \,,
	\\
	&{\rm vectormultiplet}~:
	\Psi_+, A_{2,3,4}, Y_i \,,
	\label{4d3dsplit}
	\fe
	where we have split the gaugino $\Psi$ into 
	\ie
	\Psi_\pm \equiv  {1\over 2}(1_{16}\pm \Gamma_{1890})\Psi\,,
	\fe
	and the six scalar fields $\Phi_I$ as
	\ie\label{eq:XiYidef}
	X_i=(\Phi_8,\Phi_9,\Phi_0 ),\quad Y_{j}=(\Phi_5,\Phi_6,\Phi_7 )\,.
	\fe  
	As explained in \cite{Gaiotto:2008sa}, general BPS boundary conditions for the SYM on the half space $x_1>0$ are obtained by supersymmetric configurations of the 3d multiplets in \eqref{4d3dsplit}. The generalization to interface defects is immediate via the (un)folding trick. Below we briefly review the relevant BPS boundary conditions.
	
	One simple choice involves assigning Dirichlet boundary condition for the 3d vectormultiplet and Neumann-like boundary condition for the 3d hypermultiplet
	\ie
	&{\rm D5}:~& \left.  F_{\m\n} \right|_{x_1=0} =  D_1 X_i - {1\over 2} \ep_{ijk}[X_j,X_k\left.]\right|_{x_1=0}=  \left. Y_i  \right|_{x_1=0}=0,\quad  \left. \Psi_+  \right|_{x_1=0}=0\,.
	\label{D5bc}
	\fe
	This is realized by D5-branes along the 234890 directions, intersecting with the D3-branes that lie along the 1234 directions in the 10d spacetime of type IIB string theory. Thus we refer to such boundary condition as the D5-type (or generalized Dirichlet). This is also known as the Nahm (pole) boundary condition for the $\cN=4$ SYM since the second equation in \eqref{D5bc} is the Nahm equation for $\vec X$.  Near the boundary $x_1=0$, the solutions to the Nahm equation are given by
	\ie
	X_i = - {t_i\over x_1} + {\rm regular~terms}\,,
	\fe
	for $t_i \in \mf{g}$  obeying the $\mf{su}(2)$ commutation rules
	\ie
	{}[t_i,t_j]=\ep_{ijk}t_k\,.
	\fe
	Up to a gauge transformation, $t_i$ is specified by a homomorphism $\rho:\mf{su}(2)\to \mf{g}$.
	For $\mf{g}=\mf{u}(N)$, such a homomorphism is labelled by a  partition $d=[p_1,\dots,p_k]$ of $N$ with $p_1\geq p_2\geq \dots \geq p_k>0$.  
	Correspondingly $t_i$ takes the form of a block diagonal $N\times N$ matrix
	\ie
	t_i=t^{p_1\times p_1}_i\oplus t^{p_2\times p_2}_i \oplus \cdots \oplus t^{p_k\times p_k}_i\,,
	\fe
	where each triplet $t^{p_i\times p_i}_i$ with $i=1,2,3$ gives rise to a $p_i$-dimensional irreducible representation of $\mf{su}(2)$ and explicitly
	\ie
	t_3^{p\times p}=-{i\over 2}{\rm Diag}[p-1, p-3,\dots, 1-p]\,.
	\fe 
	In the D3-D5 setup, the $U(N)$ Nahm pole labelled by the partition $d=[p_1,\dots,p_k]$ involves $k$ D5-branes and there are $p_i$ D3-branes ending on each of the $i$-th D5-branes. The trivial Nahm pole involves $N$ D5-branes and corresponds to the \textit{minimal} partition $d=[1,1,\dots,1]$ with $t_i=0$. This is the familiar Dirichlet boundary condition. The Nahm pole produced by a single D5-brane is of the \textit{regular}(principal) type with $d=[N]$.

	Another simple half-BPS boundary condition for SYM is defined by  assigning Dirichlet boundary condition for the 3d hypermultiplet and Neumann  boundary condition for the 3d vectormultiplet in \eqref{4d3dsplit}
	\ie
	{\rm NS5}:~&F_{1\m}  \left.  \right|_{x_1=0} =   X_i \left.  \right|_{x_1=0}=D_1 Y_i  \left.\right|_{x_1=0}=0,\quad \Psi_-= 0\,.
	\label{NS5bc}
	\fe
	The is realized by an NS5-brane along the 234567 directions and thus we refer to it as the NS5-type boundary condition or simply as the Neumann boundary condition for the   SYM.
	Importantly,  the D5-type and NS5-type boundary conditions are related by S-duality  in the SYM \cite{Gaiotto:2008ak}.
	
	The D5-type \eqref{D5bc} and NS5-type \eqref{NS5bc} boundary conditions can be generalized by introduce partial \textit{gauge symmetry breaking} \cite{Gaiotto:2008sa}. For a subgroup of the gauge group $H\subset G$ (which may not be simple), the corresponding Lie algebra  decomposes as
	\ie
	\mf{g} =\mf{h}\oplus \mf{h}^\perp 
	\fe 
	into the Lie algebra of $H$ and its orthogonal complement (which may not a Lie algebra). Then one can consider a mixture of NS5-type boundary condition  \eqref{NS5bc} for the components of the SYM fields in $\mf{h}$ and D5-type boundary condition \eqref{D5bc} for the components in $\mf{h}^\perp$. This defines the symmetry breaking boundary  condition associate to the subgroup $H\subset G$ \cite{Gaiotto:2008sa}.

	Finally, the half-BPS boundary conditions defined above lead  to interface defects preserving the same supersymmetry thanks to the (un)folding trick, generated by a
	$\mZ_2$-automorphism $\iota_{\rm fold}$ of the superconformal algebra $\mf{psu}(2,2|4)$ \cite{Gaiotto:2008sa}
	\ie
	\iota_{\rm fold}:~(x_1,x_2,x_3,x_4) \to (-x_1,x_2,x_3,x_4),\quad (\vec X,\vec Y )\to (-\vec X,\vec Y)\,.
	\label{folding}
	\fe
	In particular the BPS interfaces in the $\cN=4$ SYM with gauge group $G_1$ on one side $x_1>0$ and another gauge group $G_2$ on the other side $x_1<0$ correspond to BPS boundary conditions for the SYM with gauge group $G_1\times G_2$. For example, the transparent interface in the $G$ SYM corresponds to unfolding the $G\times G$ SYM with the partial symmetry breaking boundary condition that preserves the diagonal subgroup $G_{\rm diag}\subset G\times G$. Note that the flipping of $\vec X$ in \eqref{folding} ensures the continuity of the $\cN=4$ vector multiplet across the interface.

	\subsection{D5-brane interface defect and its S-dual}
\begin{figure}[!htb]
		\centering
		\begin{minipage}[t]{.48\textwidth}
			\centering
			\includegraphics[scale=2]{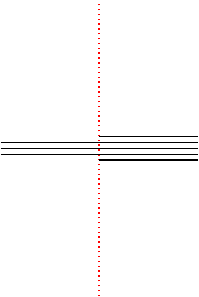}
			\subcaption{D5-brane interface} 	\label{fig:D5}
		\end{minipage}
		\hfill
		\begin{minipage}[t]{.48\textwidth}
			\centering
			\includegraphics[scale=2]{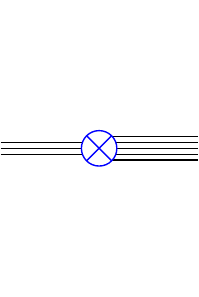}
			\subcaption{NS5-brane interface} \label{fig:NS5}
		\end{minipage}  
		\label{fig:interface}
		\caption{Half-BPS interfaces between $U(N)$ and $U(N+k)$ SYMs engineered by a single D5- or NS5-brane, related by S-duality. Here $N=3$ and $k=2$ in the figures above.  The horizontal direction is $x_1$. The solid black lines denote D3-branes in the 1234 directions. The red dotted vertical line denotes a D5-brane extending in the 234890 directions. The blue circled cross labels an NS5-brane extending in the 234567 directions. }
	\end{figure}

	We now focus on the interface interpolating between the $U(N)$ and $U(N+k)$ SYMs engineered by a single D5 brane with $N$ D3-branes on the left and $N+k$ D3 branes on the right (see Figure~\ref{fig:D5}).\footnote{We emphasize that unlike in the case without interface defects, here the $U(1)$ part of the $U(N)$ SYM does not decouple.}
	
	According to the general discussion in Section~\ref{sec:gendef}, this corresponds   by the (un)folding trick to the symmetry breaking BPS boundary condition of the $U(N)\times U(N+k)$ SYM
	\ie
	U(N)\times U(N+k)\supset U(N)_{\rm diag}
	\fe
	specified by NS5-type boundary condition  \eqref{NS5bc} for the components of the SYM fields in $\mf{ u}(N)_{\rm diag}$ and D5-type boundary condition  \eqref{D5bc} for the orthogonal complements. Consequently we have near the interface $x_1=0$, the following relations between the $U(N+k)$ fields $A_M^+$  and the $ U(N)$ fields $A_M^-$, 
	\ie
	A_\m^{+}
	=\begin{pmatrix} 
		A_\m^-	 &  *
		\\
		* & *
	\end{pmatrix},\quad 
	\vec Y^+
	=\begin{pmatrix} 
		\vec Y^- &  *
		\\
		* & *
	\end{pmatrix},\quad 
	\vec X^+
	=\begin{pmatrix} 
		\vec X^- &  *
		\\
		* & -{1\over x_1}{\vec t  }~
	\end{pmatrix}\,.
	\label{nahmpolecont}
	\fe
	Here the * labels the components of $A_M^+$ that are unconstrained (regular) at the interface. Note that due to the Nahm equation in \eqref{D5bc}, the bottom $k\times k$ block of $\vec X^+$ is given by a triplet $\vec t$ with 
	\ie
	t_3 =-{i\over 2}{\rm Diag}[k-1, k-3,\dots, 1-k]\,.
	\fe 
	
\begin{figure}[!htb]
		\begin{minipage}[c] {.25\linewidth}
			$T[SU(N)]$
		\end{minipage}	\begin{minipage}[c]{1\columnwidth}	\includegraphics[scale=1]{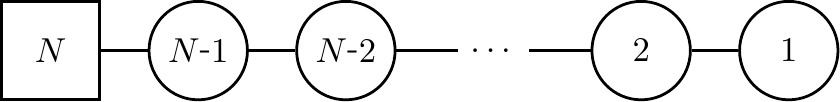}
	\end{minipage} 
	\begin{minipage}{1\linewidth}
		$$~$$
	\end{minipage}  
	\begin{minipage}[c]{.25\linewidth}
		$
		T_{[k,1,\dots,1]}[SU(N+k)]$
	\end{minipage}			
	\begin{minipage}{1\columnwidth}
		\includegraphics[scale=1]{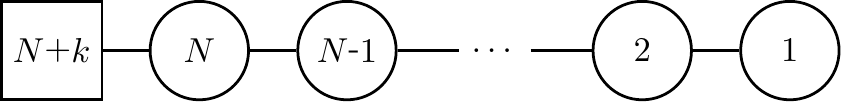}
	\end{minipage}  
	\begin{minipage}{1\linewidth}
		$$~$$
	\end{minipage}  
	\begin{minipage}{.12\linewidth}
		D5-brane Interface 
	\end{minipage}	 	 \begin{minipage}[c]{1\columnwidth}
	\includegraphics[scale=1]{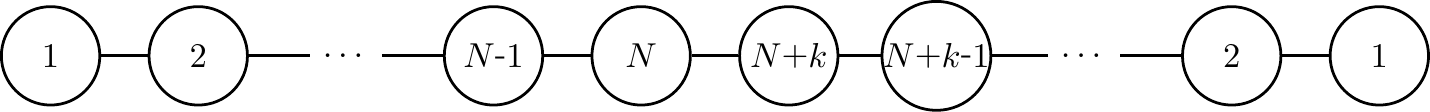}
\end{minipage}
\caption{The ultraviolet linear quiver descriptions for the 3d $\cN=4$ SCFTs $T[SU(N)]$ and $T[SU(N+k)]$, as well as (the mirror dual for) the D5-brane interface theory. Each numbered circle  node denotes a 3d $\cN=4$ vector multiplet for the correspondence unitary gauge group, and each segment denotes a bifundamental hypermultiplet. The boxed nodes are not gauged and carry the Higgs branch flavor symmetries of the corresponding SCFTs.  }
\label{fig:quiver}
\end{figure}

	The S-dual of the D5 brane interface in the $\cN=4$ SYM with complexified gauge coupling $\tau={4\pi i \over g_4^2}+{\theta \over 2\pi}$ is easy to infer from the IIB brane setup. By an S-duality transformation of the IIB string theory, the D5 brane interface  maps to an NS5 brane interface in Figure~\ref{fig:NS5}. The resulting 4d theory on the D3-branes  consists of two individual 4d SYMs with $U(N+k)$ and $U(N)$ gauge  groups and Neumann boundary conditions, coupled together by a bifundamental hypermultiplet localized at the interface. The SYM gauge couplings are now at the S-dual value $\tau'=-{1\over \tau}$. By performing a further S-duality transformation on each of the two SYM factors, the $U(N)$ SYM with coupling $\tau'$ and Neumann boundary condition maps to the $U(N)$ SYM at original gauge coupling $\tau$ coupled to a three-dimensional $\cN=4$ SCFT known as the $T[SU(N)]$ theory, similarly for the $U(N+k)$ factor where S-duality introduces the coupling to the $T[SU(N+k)]$ SCFT (see Figure~\ref{fig:quiver} for the quiver description  of these 3d theories). In general the 3d $T[G]$ SCFT has global symmetries $G\times G^\vee$ where $G^\vee$ denotes the Langlands dual of the 4d gauge group. Here the global symmetries of the $T[SU(N)]$ and $T[SU(N+k)]$ theories are gauged by the $SU(N)$ and $SU(N+k)$ vector multiplets in the bulk, as well as on the interface along with the bifundamental hypermultiplet.  
	
	Note that $T[SU(N)]$ coupled to the $U(N)\times U(N+k)$ bifundamental multiplet defines the $T_{[k,1,\dots,1]}[SU(N+k)]$ SCFT with $SU(N)\times SU(N+k)$ global symmetry. This belongs to a generalization of the $T[G]$ SCFT by two homomorphisms $\rho$ and $\sigma$ from $\mf{su}(2)$ to $\mf{g}$, labelled as $T_\rho^\sigma[G]$. When either $\rho$ or $\sigma$ is the trivial homomorphism, we drop the corresponding sub- or super-script. In particular the $T[G]$ theory corresponds to trivial $\rho$ and $\sigma$. The $T_\rho^\sigma[G]$ SCFT is engineered by considering 4d $\cN=4$ $G$ SYM on a segment (suppressing the non-compact $\mR^3$) with Nahm pole boundary condition  for $\vec X$ labelled by $\rho$ at one end, and Nahm pole boundary condition for $\vec Y$ labelled by $\sigma$ at the other end, together with an S-duality wall in the middle \cite{Gaiotto:2008ak}. By gluing with the 4d $G$ SYM on a half-space with suitable boundary conditions, the coupling to $T_\rho^\sigma[G]$ SCFT induces the S-dual boundary condition for the 4d SYM.\footnote{See \cite{Dedushenko:2018tgx} for recent work on the factorization and gluing of supersymmetric partition functions.} For this reason, the $T_\rho^\sigma[G]$ SCFT is sometimes referred to as the S-duality transformation kernel. 
	Back to the case we are interested in here, which is due to a single nontrivial Nahm pole $\rho=[k,1,1,\dots,1]$. The relevant 3d defect theory is given by $T_{[k,1,1,\dots,1]}[SU(N+k)]$ (see Figure~\ref{fig:quiver}).

\subsection{Identifying the defect-Yang-Mills sector}
\label{sec:dYM}
The D5 brane interface preserves the half-BPS algebra $\mf{osp}(4|4)$. As explained in \cite{Wang:2020seq}, this subalgebra  contains the 4d supercharge $\cQ$ of \cite{Pestun:2009nn} which squares to a combination of the transverse rotation $M_\perp$ to an two-sphere
\ie
S^2_{\rm YM}:~x_4=0,~ \sum_{i=1}^3 x_i^2=1\,,
\fe
and the R-symmetry generator $R_{56}$ inside $\mf{so}(6)_{R}$.\footnote{We follow the convention of \cite{Wang:2020seq} for the generators of the superconformal algebra $\mf{psu}(2,2|4)$. In particular the $\mf{so}(6)_R$ symmetry is generated by $R_{IJ}$ with $I,J=5,6,\dots,9,0$.} Furthermore, after mapping the setup to $S^4$ by the stereographic projection, one can carry out the $\cQ$-localization of the 4d SYM with a general half-BPS interface defect. The dynamics of general defect observables in the $\cQ$ cohomology is determined by an effective 2d defect-Yang-Mills (dYM) theory on the $S^2_{\rm YM}$, which is described by the constrained 2d Yang-Mills \cite{Pestun:2009nn} coupled to certain topological quantum mechanics (TQM) \cite{Dedushenko:2016jxl} on the equator
\ie
S^1_{\rm TQM}:~x_4=0,~ x_1=0,~\sum_{i=2}^3 x_i^2=1\,.
\fe
The dYM gives an efficient way to determine defect observables in the $\cQ$-cohomology by standard two-dimensional gauge theory techniques, which oftentimes reduce to  computations in (multi)matrix models. 

The details of the dYM  will depend on the specific interface. This is especially subtle when the defect introduces singularities for the SYM fields, such as the Nahm pole  for the D5-brane interface. Here we will circumvent this by two ways to determine the dYM, namely the S-duality of the SYM and the mirror dual of the interface (boundary condition). As explained in \cite{Gaiotto:2008ak}, the two are closely related which we review below.

Let's first consider the S-dual configuration in Figure~\ref{fig:NS5} involving an NS5-brane interface, described by two individual SYMs on half-spaces with Neumann boundary conditions and gauge groups $U(N)$ and $U(N+k)$. The two SYM factors are coupled together just by a bifundamental hypermultiplet of $U(N)\times U(N+k)$ localized at the interface.
The invariance of the bulk SYM under S-duality transformations ensures that the correlation function of defect observables in the SYM with  D5-brane interface is equivalent to that with an NS5-brane interface. 
A particular class of observables in the $\cQ$-cohomology of the SYM are ${1\over 8}$-BPS operators on the $S^2_{\rm TQM}$ \cite{Giombi:2009ds}
\begin{shaded}
	\ie
	\cO_J(x_i)\equiv \tr(x_1\Phi_7+x_2\Phi_9+x_3\Phi_0+i\Phi_8)^J\,.
	\label{OJ}\fe
\end{shaded}
\noindent
Since these operators (individually half-BPS) are invariant under S-duality \cite{Intriligator:1998ig}, their correlators in the presence of D5- or NS5-brane interface are simply related by
\ie
\la \cO_{J_1} \cdots \cO_{J_n}\ra_{\rm D5}^{\tau} \equiv \la \cO_{J_1} \cdots \cO_{J_n}\ra_{\rm NS5}^{\tau'}\,,
\fe
where $\tau'=-{1\over \tau}$.

In a later section, we will derive a matrix model for the one point function of the D5-brane interface $\la \cO_{J}\ra_{\rm D5}$. For this purpose, it's convenient to use the S-dual NS5-brane description, which leads to a simple dYM description which we now present.

Following the general discussion in \cite{Wang:2020seq}, the $\cQ$-cohomology is parametrized by emergent 2d YM gauge fields $\cA'$ and $\cA$ on the two hemispheres $HS^2_\pm$ with gauge group $U(N+k)$ and $U(N)$ respectively, as well as the (twisted combination of) hypermultiplet scalars  $(Q,\tilde Q)$ in the bifundamental representation of $U(N)\times U(N+k)$.
The partition function of the  dYM  is
\ie
Z_{\rm dYM}=\int_{S^2_{\rm YM}} D \cA |_{HS^2_-} D \cA'|_{HS^2_+} D Q D\tilde Q |_{S^1_{\rm TQM}} e^{-S_{\rm YM}(\cA)-S_{\rm YM}(\cA')-S_{\rm TQM}(\cA,\cA',Q,\tilde Q)}\,.
\label{ZdYM}
\fe
Here the YM action $S_{\rm YM}(\cA)$ is
\ie
S_{\rm YM}(\cA) \equiv -{ 1\over   g_{\rm YM}^2}\int_{HS^2_-}   dV_{S^2} \tr (\star \cF)^2,
\quad 
g_{\rm YM}^2=-{8\pi \over g_4^2 R^2}\,,
\label{YM2}
\fe
with the field strength $\cF\equiv d\cA+\cA \wedge \cA$ and similarly for $S_{\rm YM}(\cA')$ on $HS^2_+$. The TQM  is described by
\ie
S_{\rm TQM}=&-2\pi \int d\varphi \, \tilde Q_i^m (D_{\cA,\cA'})^{i n}_{j m} Q^j_n\,,
\fe
where $i,j=1,\dots,N$ and $m,n=1,\dots,N+k$ are  indices for $U(N)$ and $U(N+k)$ fundamental representations respectively. The gauge covariant derivative is
\ie
(D_{\cA,\cA'})^{i n}_{j m}=d\D^i_j \D^n_m + \D^n_m\cA^i{}_j+\D^i_j\cA'_m{}^n \,.
\fe
Integrating out the 2d/1d fields in \eqref{ZdYM} as described in \cite{Wang:2020seq}, we obtain a (two-)matrix model 
\ie
&Z_{\rm D5}=   \int  [da][d\A]
\underbrace{\Delta(a)  \bD(a) e^{-{g_4^2\over 4} \sum_{i=1}^N a_i^2 }  \over N! }_{U(N)~{\rm SYM~Neumann~b.c.}}
\underbrace{\Delta(\A )  \bD(\A) e^{-   {g_4^2\over 4}  \sum_{n=1}^{N+k} \A_n^2 }  \over (N+k)! }_{U(N+k)~{\rm SYM~Neumann~b.c.}}
\underbrace{
	1
	\over  \prod_{i=1}^N \prod_{n=1}^{N+k} 2\cosh(\pi (a_i+\A_n))}_{U(N)\times U(N+k)~{\rm bifundamental~hyper}}\,.
\label{NS5mm}
\fe
with measure
\ie
{}[da]\equiv\left(\prod_{n=1}^N da_j\right),\quad [d\A]\equiv \left(\prod_{n=1}^{N+k} d \A_n\right) \,.
\fe
Here $\Delta(\cdot)$ denotes the usual Vandermonde determinant
\ie
{  \Delta} (z) \equiv \prod_{i<j}  (z_i-z_j)\,,
\fe
while $\bD(\cdot)$ denotes the one-loop determinant for 3d $\cN=4$ vector multiplets
\ie
{\bm \Delta} (z) \equiv \prod_{i<j} 2\sinh \pi (z_i-z_j)\,.
\fe
In writing \eqref{NS5mm}, we have made manifest its factorized form, from gluing  $U(N)$ and $U(N+M)$ SYMs with Neumann boundary conditions and coupled together by the bifundamental hypermultiplet. 
Finally, the one point function  of the ${1\over 8}$-BPS operator   $\cO_J$ on the hemispheres $HS^2_{\pm}$ simply amounts to an insertion  of   $\sum_{n=1}^{N+k} (\A_n)^J$ or $\sum_{i=1}^N (a_i)^J$ respectively in \eqref{NS5mm} \cite{Wang:2020seq}.\footnote{As in the case without the interface, the correlation functions of the local operators $\cO_J$ are topological \cite{Giombi:2012ep} and thus independent of the location of the insertions on the $S^2_{\rm YM}$ as long as it doesn't cross the interface at $S^1_{\rm TQM}$ \cite{Wang:2020seq}. } 
\begin{figure}[!htb]
	\centering
	\begin{minipage}[t]{.35\textwidth}
		\centering
		\includegraphics[scale=3]{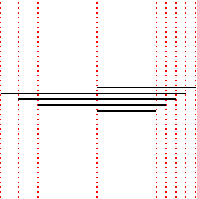}
		\subcaption{Decouple 4d vector multiplets} 
	\end{minipage}
	\hfill
	\begin{minipage}[t]{.50\textwidth}
		\centering
		\scalebox{ -1}[1]{ \includegraphics[scale=2]{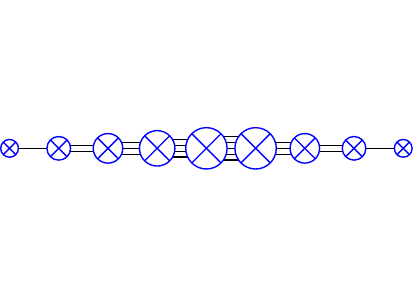} }
		\subcaption{S-dual brane configuration} 
	\end{minipage}  
	\caption{The standard procedure to identify the mirror dual for the D5-brane interface. In the left figure, we ungauge the 4d vector multiplets by attaching the individual D3 branes to D5 branes, implementing Dirichlet boundary conditions. By performing an S-duality transformation, we obtain the brane configuration in the right figure, which gives the mirror dual quiver description of the D5-brane interface.}
	\label{fig:mirror}\end{figure}

In a sense, the dYM  presented above from the NS5-brane interface is S-dual to the would-be dYM that directly descends from the D5-brane description which we do not pursue here directly due to the singular Nahm pole boundary condition. However it is known that such superconformal boundary condition has an ultraviolet (UV) description  by certain $\cN=4$ supersymmetric linear quiver gauge theories \cite{Gaiotto:2008ak}. This UV Lagrangian is   mirror dual to the original interface theory (boundary condition) since it couples to the bulk $\cN=4$ vector multiplets via the 3d moment map operators on the Coulomb branch, as opposed to the Higgs branch.  There is a standard procedure to identify the mirror dual Lagrangian as explained in  \cite{Gaiotto:2008ak}. This follows from first isolating the interface degrees of freedom and bulk-interface couplings by introducing D5 branes that impose Dirichlet boundary conditions on the 4d vector multiplets away from the interface, and then implementing the IIB S-duality on this configuration.
As reviewed in Figure~\ref{fig:mirror}, carrying out this procedure, the D5-brane interface between $U(N)$ and $U(N+k)$ SYMs here has a UV Lagrangian description given by the bottom quiver in Figure~\ref{fig:quiver}. Following \cite{Wang:2020seq}, one can then localize the coupled system with this Lagrangian interface theory and obtain a dYM similar to that in \eqref{ZdYM}. The main differences are that the 2d Yang-Mills coupling becomes 
\ie
g_{\rm YM}^2=-{g_4^2\over 2\pi  R^2}\,,
\fe
and the TQM is now an interacting theory descending from the D5-interface quiver in Figure~\ref{fig:quiver}. We will not need the explicit form of this dYM description in this paper but we point out that it leads to a different form of the integrand for the matrix model than in \eqref{NS5mm}, though the integrals agree as expected (see also the next section for explicit formulae).\footnote{We also emphasize that the D5-interface quiver here is of the \textit{bad} type according to the \textit{good-bad-ugly} classification in \cite{Gaiotto:2008ak}. This means as a stand-alone 3d $\cN=4$  theory, it is obscure from the quiver what the correct IR description should be (e.g. there are monopole operators that violate the putative unitarity bound, spontaneous symmetry breaking etc.). Nevertheless as we will see in the next section, this bad quiver suffices in describing the D5-brane interface defect, at least at the the level of the $S^3$ partition function (which contributes to the interface matrix model). }

\subsection{Simplified single-matrix model for D5-brane interface and Nahm pole} \label{sec:simmm}
In this section we simplify the matrix model \eqref{NS5mm} by explicitly integrating out the $U(N+k)$ degrees of freedom. The final result is 
\begin{shaded} \ie 
	Z_{\rm D5} =&
	{ C_{N,k} \over N!}\int [da]{\Delta(a)^2 \prod_{i=1}^N \prod_{j=1}^k (a_{i}-i{2j-k-1\over 2})	  \over
		\prod_{j=1}^N 2\cosh {\pi (a_j+{ki\over 2})} 
	}
	e^{ - {8\pi^2\over g_4^2} \sum_{i=1}^N a_i^2}\,,
	\label{D5mmsim}
	\fe
	with
	\ie
	C_{N,k}=   \left({4\pi \over g_4^2}\right)^{N^2+(N+k)^2\over 2} e^{{4\pi \over g_4^2}{ k(k-1)(k+1)\over 24}} (-i)^{Nk }G(1+k)\,,
	\fe
	where $G(z)$ is the Barnes G-function.
\end{shaded}

There are several ways to derive   \eqref{D5mmsim},  relying on a generalized Cauchy determinant formula and repeated Fourier transformations. We present one derivation below which makes clear the connections to the descriptions of the D5-brane interface by S-duality and mirror symmetry as explained in the previous section.

We start by rewriting \eqref{NS5mm} using the S-duality kernel $Z^{T[SU(N)]}(a,a')$ which acts on the wavefunction of the $U(N)$ SYM on $HS^4$ with Dirichlet boundary condition (after dropping an unimportant phase factor compared to \cite{Wang:2020seq}) 
\ie
Z^{\rm SYM}_{\rm Dir}(a, g_4^2 )
=
e^{ -{4\pi^2\over g_4^2}   \sum_{i=1}^N a_i^2 }
{ \Delta(a )\over   \bD(a )  }\,,
\fe
as
\ie
{1\over N!}\int  [da'] {\bm \Delta}^2(a')Z^{T[SU(N)]} (a', a) Z^{\rm SYM}_{\rm Dir}(a', g_4^2)=\left({g_4^2\over 4\pi }\right)^{N^2\over 2}  Z^{\rm SYM}_{\rm Dir}\left(a, { 16 \pi^2 \over  g_4^2 }\right)\,.
\fe
More explicitly, the S-duality kernel for the $U(N)$ SYM corresponds to the partition function of the 3d $T[SU(N)]$ SCFT  \cite{Benvenuti:2011ga,Nishioka:2011dq}
\ie
Z^{T[SU(N)]} (m_i,e_j)={\sum_{\rho \in S_N}{  (-1)^{|\rho|} e^{2\pi i \sum_j^N m_{\rho(j)}  e_j }  } \over  i^{N(N-1)\over 2}  \bD(m) \bD(e)}\,,
\label{TSUN}
\fe
where $m_i$ and $e_j$ 
are the  mass parameters (FI parameters) for the $SU(N)\times PSU(N)$ global symmetries respectively.\footnote{The mass parameters are subjected to the constraints $\sum_i m_i=\sum_i e_i=0$.}   
Similarly the partition function  $Z^{T[SU(N+k)]}(\A,\A')$ for the $T[SU(N+k)]$ SCFT defines the S-duality kernel for the $U(N+k)$ SYM.  

Consequently the SYM partition function with D5-brane interface \eqref{NS5mm} becomes\footnote{In this section, we  use $a,a'$ and $\A,\A'$ to denote eigenvalues of hermitian matrices of rank $N$ and $N+k$ respectively.}
\ie
& Z_{\rm D5}=
\left({4\pi \over g_4^2}\right)^{N^2+(N+k)^2\over 2}
\int 
[da] [d\A][da'] [d\A'] \underbrace{\bD(a)^2\bD(a')^2\bD(\A)^2\bD(\A')^2\over (N!)^2((N+k)!)^2}_{3d~{\rm vector multiplet}}
\underbrace{\Delta(a)   e^{-{ 4 \pi^2 \over  g_4^2 }   \sum_{i=1}^N a_i^2 }  \over   \bD(a) }_{U(N)~{\rm SYM~Dirichlet~b.c.}}
\\
&
Z^{T[SU(N)]}(a,a')\underbrace{
	1
	\over  \prod_{i=1}^N \prod_{n=1}^{N+k} 2\cosh(\pi (a'_i+\A'_n))}_{U(N)\times U(N+k)~{\rm bifundamental~hyper}}
Z^{T[SU(N+k)]}(\A,\A') \underbrace{\Delta(\A)    e^{ -{ 4 \pi^2 \over  g_4^2 }  \sum_{n=1}^{N+k} \A_n^2 }  \over   \bD(\A)}_{U(N+k)~{\rm SYM~Dirichlet~b.c.}}\,.
\label{D5mm}
\fe
From the quiver descriptions of the 3d $\cN=4$ theories in Figure~\ref{fig:quiver}, we note the following relation between their  $S^3$ partition functions.  The $T_{[k,1,\dots,1]}[SU(N+k)]$ theory can be obtained from gauging the $U(N)$ symmetry of $T[SU(N)]$ and $N+k$ fundamental $U(N)$ hypermultiplets,\footnote{The global symmetry of the  $T_{[k,1,\dots,1]}[SU(N+k)]$ theory is $U(N)\times SU(N+k)$. Consequently, the partition function $Z^{T_{[k,1,\dots,1]}[SU(N+k)]}(a,\A )$ depends on the overall shifts of $a_i$, unlike in the case of $T[SU(N)]$.}
\ie
Z^{T_{[k,1,\dots,1]}[SU(N+k)]}(a,\A )={1\over N!}\int [da']  \bD(a')^2 
{Z^{T[SU(N)]}(a,a')
	\over  \prod_{i=1}^N \prod_{n=1}^{N+k} 2\cosh(\pi (a'_i+\A_n))} \,.
\label{gluingTSU}
\fe
The mirror quiver of the D5-brane interface theory is obtained from gauging the diagonal $SU(N+k)$ symmetry of $T_{[k,1,\dots,1]}[SU(N+k)]$ and $T[SU(N+k)]$,
\ie
K(a,\A)={1\over (N+k)!}\int [d\A'] \bD(\A')^2  Z^{T_{[k,1,\dots,1]}[SU(N+k)]}(a,\A')Z^{T[SU(N+k)]}(\A,\A')\,.
\label{gluingK}
\fe
Therefore, the matrix model in \eqref{D5mm} can be rewritten as 
\ie
& Z_{\rm D5}=
{1\over N! (N+k)!} \int [da][d \A]  \bD(a)^2\bD(\A)^2 Z_{\rm Dir}(a,g_4^2) K(a,\A)   Z_{\rm Dir}(\A,g_4^2) 
\label{D5mmKt}
\fe
in accordance with the second dYM description given in the last section, where the D5-brane interface (before S-duality) is described  by coupling to a 3d $\cN=4$ quiver theory in the UV with partition function $K(a,\A)$. Below we will derive an explicit form for $K(a,\A)$ which leads to \eqref{D5mmsim}.

\subsubsection*{Partition function of $\bm {T_{[k,1,\dots,1]}[SU(N+k)]}$}
We start by evaluating the partition function for $T_{[k,1,\dots,1]}[SU(N+k)]$. From \eqref{gluingTSU}, we have 
\ie
Z^{T_{[k,1,\dots,1]}[SU(N+k)]}(a,\A)=&{(-i)^{{N(N-1)\over 2}}\over  N! \bD(a)}\int  [d a']
\sum_{\sigma\in S_N} (-1)^\sigma e^{2\pi i \sum_{j=1}^N a'_{\sigma(j)} a_{j} }
\\
&\times
{ 
	\bD(a')
	\over
	\prod_{i}^N\prod_{n=1}^{N+k} 2\cosh(\pi(a'_i+\A_n))
}\,,
\label{ZTk1}
\fe
where we do not impose $\sum_i a'_i=0$ or  $\sum_i a_i=0$ in the integrand.\footnote{The $T[SU(N)]$ sector only couples to the traceless part of $a'$, which can be achieved (without the constraint $\sum_i a'_i=0$) by shifting $a'_i \to a'_i-a'_N$ which does not modify the $T[SU(N)]$ partition function as long as $\sum_i a_i=0$. Meanwhile the trace part of $a'$ couples nontrivially to the hypermultiplets and allows for an FI term $e^{2\pi i \eta_N \sum_{i=1}^N a'_i}$ in the partition function \eqref{ZTk1}, which is equivalent to further relaxing the constraint $\sum_i a_i=0$.
}  The SCFT has $U(N) \times  SU(N+k)$ global symmetry with mass parameters given by $a_i$ and $\A_m$.

Recall the generalized Cauchy determinant formula in \cite{Honda:2013pea} (without constraints on $a$ and $\A$)
\ie
{ 
	(-1)^{N(N-1)\over 2}	e^{  \pi k (\sum_{i=1}^N a_i+\sum_{n=1}^{N+k} \A_n)}\bD(a)\bD(\A)
	\over
	\prod_{i=1}^N\prod_{n=1}^{N+k} 2\cosh(\pi(a_i+\A_n)) )
}
=
\det\bigg(
{\theta_{N,n}\over 2\cosh ({\pi(a_n+\A_m)})}
+e^{2\pi(N+k+{1\over 2}-n)\A_m} \theta_{n,N+1}
\bigg)_{m,n}\,,
\fe
where
\ie
\theta_{mn}\equiv \begin{cases} 1 &m\geq n \,,\\ 0 &n<m \,.\end{cases}
\fe
This allows us to simplify \eqref{ZTk1} 
\ie
&Z^{T_{[k,1,\dots,1]}[SU(N+k)]}(a,\A)
\\=&{i^{N(N-1)\over 2}  \over N!\bD(a) \bD(\A)}\int [da']
\,  e^{ - \pi k  \sum_{i=1}^N a'_i } 
\sum_{\sigma\in S_N} (-1)^{|\sigma|} e^{2\pi i \sum_{j=1}^N a'_{\sigma(j)} a_{j} }
\\
&\times
\sum_{\rho\in S_{N+k}} (-1)^{|\rho|} \prod_{m=1}^{N+k} \bigg(
{\theta_{N,m}\over 2\cosh {\pi(\A_{\rho(m)}+a'_m)}}
+e^{2\pi(N+k+{1\over 2}-m)\A_{\rho(m)}} \theta_{m,N+1}
\bigg) \,,
\\
=&  {i^{N(N-1)\over 2}  \over N!\bD(a) \bD(\A)}\int [da']
\, e^{-\pi k  \sum_{i=1}^N a'_i}
e^{2\pi i \sum_{j=1}^N a'_{j} a_{j} }
\\
&\times
\sum_{\sigma\in S_N}\sum_{\rho\in S_{N+k}} (-1)^{|\sigma| }(-1)^{|\rho| }\prod_{j=1}^N
{1\over 2\cosh {\pi(\A_{\rho(j)}+a'_{\sigma(j)})}}
\prod_{m=N+1}^{N+k} e^{2\pi(N+k+{1\over 2}-m)\A_{\rho(m)}}\,,
\label{ZTk2}
\fe
where in the second equality, we have made a change of variables $a_{j} \to a_{\sigma^{-1}(j)}$ and then sent $\sigma \to \sigma^{-1}$ in the sum over $S_N$ permutations.

Next we use the Fourier transformation 
\ie
{1\over  \cosh \pi y}=\int_{-\infty}^\infty dx {e^{2\pi i x y}\over  \cosh \pi x}
\fe
for each cosh factors in \eqref{ZTk2} to obtain 
\ie
&Z^{T_{[k,1,\dots,1]}[SU(N+k)]}(a,\A)
\\
=& {i^{N(N-1)\over 2}  \over N!\bD(a) \bD(\A)}\int [da'] \int \left( \prod_{i=1}^N {dx_i  \over 2\cosh {\pi x_i}} \right)
\, e^{-\pi k  \sum_{i=1}^N a'_i}
e^{2\pi i \sum_{j=1}^N a'_{j} a_{j} }
\\
&\times
\sum_{\sigma\in S_N}\sum_{\rho\in S_{N+k}} (-1)^{|\sigma| }(-1)^{|\rho| } 
e^{2\pi i \sum_{j=1}^N x_j (\A_{\rho(j)}+a'_{\sigma(j)})}  
\prod_{m=N+1}^{N+k} e^{2\pi(N+k+{1\over 2}-m)\A_{\rho(m)}}\,,
\\
=& {i^{N(N-1)\over 2} \over \bD(a) \bD(\A)}\int [da'] \int \left( \prod_{i=1}^N {dx_i  \over 2\cosh {\pi x_i}} \right)
\, e^{-\pi k  \sum_{i=1}^N a'_i}
e^{2\pi i \sum_{j=1}^N a'_{j} a_{j} }
\\
&\times
\sum_{\rho\in S_{N+k}} (-1)^{|\rho| } 
e^{2\pi i \sum_{j=1}^N x_j (\A_{\rho(j)}+a'_{ j})}  
\prod_{m=N+1}^{N+k} e^{2\pi(N+k+{1\over 2}-m)\A_{\rho(m)}}\,,
\fe
after a change of variables $x_j\to x_{\sigma(j)}$ and redefining the $S_{N+k}$ permutation as $\rho \to \rho \cdot \sigma $ in the second equality. The final step is to integrate out the $U(N)$ variables $a'_i$, leading to delta functions
\ie
\prod_{j=1}^N \D (a_j +{ki\over 2}+x_j)\,,
\fe
and doing the $x_i$ integrals gives\footnote{The $S^3$ partition function for general $T_\rho^\sigma[G]$ theories was conjectured and proven for the case $G=SU(N)$ and $\sigma=[1,\dots,1]$ in \cite{Nishioka:2011dq} by induction.}
\begin{shaded}
	\ie
	&Z^{T_{[k,1,\dots,1]}[SU(N+k)]}(a,\A)
	= {i^{N(N-1)\over 2}  	\sum_{\rho\in S_{N+k}} (-1)^{|\rho| }   e^{ \pi \sum_{m=N+1}^{N+k}(2N+k+1-2m)\A_{\rho(m)}}	e^{-2\pi i \sum_{j=1}^N   a_j  \A_{\rho(j)} }    \over \bD(a) \bD(\A)    \prod_{j=1}^N   2\cosh {\pi (a_j +{k \over 2}i)}} \,.
	\label{ZTkmm}
	\fe
\end{shaded}

\subsubsection*{Partition function of D5-brane interface quiver theory}
Let us now study the partition function of the D5-brane interface quiver theory using \eqref{gluingK}. Plugging in the partition function \eqref{ZTkmm} for the $T_{[k,1,\dots,1]}[SU(N+k)]$ and \eqref{TSUN} for the $T[SU(N+k)]$ theories, we have
\ie
K(a,\A)
=&{i^{N(N-1)\over 2}  (-i)^{(N+k)(N+k-1)\over 2} 	\over
	(N+k)!  \bD(a)\bD(\A)
	\prod_{j=1}^N 2\cosh {\pi (a_j+{ki\over 2})}
} \int  [d\A']
\\
&
\sum_{\rho\in S_{N+k}} (-1)^{|\rho|} e^{2\pi i \sum_{m=1}^{N+k} \A'_{\rho(m)} \A_{m} }
\sum_{\sigma\in S_{N+k}}  (-1)^{|\sigma|} 
e^{-2\pi i \sum_{j=1}^N  a_j \A'_{\sigma (j)}}
e^{\pi \sum_{n=N+1}^{N+k}(2N+k+1-2n)\A'_{\sigma (n)}} \,,
\\
=&{ (-i)^{k(k-1)+2N k\over 2}	\over
	\bD(a)\bD(\A)
	\prod_{j=1}^N 2\cosh {\pi (a_j+{ki\over 2})}
} \int  [d\A']
\\
&
e^{2\pi i \sum_{m=1}^{N+k} \A'_{m} \A_{m} }
\sum_{\sigma\in S_{N+k}}  (-1)^{|\sigma|} 
e^{-2\pi i \sum_{j=1}^N  a_j \A'_{\sigma (j)}}
e^{\pi \sum_{n=N+1}^{N+k}(2N+k+1-2n)\A'_{\sigma (n)}}\,. 
\fe
In the second equality above, we have made the change of variables $\A'_m \to \A'_{\rho^{-1}(m)}$ and then redefined the permutation $\sigma \to \rho \cdot \sigma$. Now integrating out the $U(N+k)$ eigenvalues $\A'_m$, we get

\begin{shaded}
	\ie
	K(a,\A)
	=&{ (-i)^{k(k-1)+2N k\over 2} \sum_{\sigma\in S_{N+k}}  (-1)^\sigma 
		\prod_{m=1}^{N+k }\D(\A_m-\xi_{\sigma(m)}) 	\over
		\bD(a)\bD(\A)
		\prod_{j=1}^N 2\cosh {\pi (a_j+{ki\over 2})}
	}
	\label{Kpf}
	\fe
	\noindent where  
	\ie
	\xi_m=(a_1 ,a_2 ,\dots, a_N ,i{k-1\over 2},i{k-3\over 2},\dots,i{1-k\over 2} )\,.
	\fe
\end{shaded}\noindent
Note that the presence of the delta functions in \eqref{Kpf} is a feature of the \textit{bad} quiver which is essential here to provide the expected continuity condition \eqref{nahmpolecont} for the ${\mf{u}}(N)$ components of the SYM fields. 

Plugging the expression \eqref{Kpf} for $K(a,\A)$ into \eqref{D5mmKt} together with the hemisphere wavefunctions of the $U(N)$ and $U(N+k)$ SYMs with Dirichlet boundary conditions, and integrating out the delta functions, would give the single matrix model in \eqref{D5mmsim} for the D5-brane interface. Below we do it in two steps.

\subsubsection*{Hemisphere matrix model with Nahm pole boundary condition}
As a by-product and also an intermediate step, we can determine the  hemisphere wavefunction  of the $U(N+k)$ SYM with the Nahm pole boundary condition labelled by $\rho=[k,1,\dots,1]$. The IR superconformal boundary condition arises from coupling the SYM on half-space (hemisphere) to the mirror quiver theory (last one in Figure~\ref{fig:quiver}) in the UV. The supersymmetric wavefunction for the Nahm pole is then given by the following matrix model,
\ie
Z_{\rm Nahm}(a,g_4^2) =& {1\over (N+k)!} \int [d \A]  \bD(\A)^2  K(a,\A)  Z_{\rm Dir}(\A, g_4^2)
\\
=& (-i)^{Nk } {  \Delta(a) \prod_{j=1}^N \prod_{n=1}^k (a_{j} -i{2n-k-1\over 2})	\prod_{1\leq m<n \leq k}  (n-m) \over
	\bD(a)	\prod_{j=1}^N 2\cosh {\pi (a_j+{ki\over 2})} \,,
}
\\
&\times 
e^{  -{ 4 \pi^2 \over  g_4^2 }   \sum_{i=1}^N a_i^2} e^{  { 4 \pi^2 \over  g_4^2 }    { k(k-1)(k+1)\over 24}  }\,.
\fe
This gives
\begin{shaded}
	\ie
	Z_{\rm Nahm}(a,g_4^2) =&
	C'_{N,k}{ \Delta(a) \prod_{j=1}^N \prod_{n=1}^k (a_{j}-i{2n-k-1\over 2})	  \over
		\bD(a)
		\prod_{j=1}^N 2\cosh {\pi (a_j+{ki\over 2})} 
	}
	e^{  -{ 4 \pi^2 \over  g_4^2 } \sum_{i=1}^N a_i^2}\,.
	\label{nahmmm}
	\fe with
	\ie
	C'_{N,k}=   e^{     { \pi^2  k(k-1)(k+1)\over 6 g_4^2}  }  (-i)^{Nk}G(1+k)\,.
	\fe
\end{shaded}
\noindent
Next the D5-brane interface matrix model is related to \eqref{nahmmm} by gluing with the hemisphere wavefunction of $U(N)$ SYM with Dirichlet boundary condition
\ie
Z_{\rm D5}=  \left ({ 4 \pi  \over  g_4^2 } \right)^{N^2+(N+k)^2\over 2} {1\over N!} \int [da] \bD(a)^2     Z_{\rm Dir}(a,g_4^2)    Z_{\rm Nahm}(a,g_4^2) \,,
\fe
which gives \eqref{D5mmsim} with $C_{N,k}= \left ({ 4 \pi /  g_4^2 } \right)^{N^2+(N+k)^2\over 2} C'_{N,k}$.

\section{Defect One-Point Function from the D5-brane Matrix Model}
\label{sec:D5onepf}

In this section, using the D5-brane interface matrix model \eqref{D5mmsim}, we compute the one-point function $\la \cO_J\ra_\cD$ of the half-BPS operators $\cO_J$ \eqref{OJ} inserted at the north pole $x_\m=(1,0,0,0)$ on $S^4$ with the D5-brane interface defect $\cD$ along the equator $S^3$ at $x_1=0$. Since we plan to compare the field theory results with IIB string theory and integrability methods, we work with the planar large $N$ limit where the usual 't Hooft coupling $\lambda =g_{4}^2 N$ is held fixed.

The $k=0$ case of the D5-brane interface is particularly simple due to the absence of a Nahm pole in \eqref{nahmpolecont}. The 4d/3d system is simply described by $\cN=4$ $U(N)$ SYM coupled to a 3d hypermultiplet in the fundamental representation at $x_1=0$. The corresponding matrix model is simply\footnote{This matrix model first appeared in \cite{Robinson:2017sup}.} 
\ie 
\cZ  =&
\int [da]{\Delta(a)^2  	  \over
	\prod_{j=1}^N 2\cosh {\pi  a_j } 
}
e^{ - {8\pi^2 N\over \lambda} \sum_{i=1}^N a_i^2}\,,
\label{k=0mm}
\fe
where we have dropped overall constants in \eqref{D5mmsim} which are irrelevant for computing the one-point functions $\la \cO_J\ra_\cD$. In \cite{Wang:2020seq} the one-point function $\la \cO_J\ra_\cD$ was computed using the above matrix model in the strong coupling limit to the first nontrivial order in the $1\over \lambda$ expansion for the $k=0$ D5-brane interface defect. The results are in perfect agreement with a probe brane analysis in IIB string theory on $AdS_5\times S^5$ \cite{DeWolfe:2001pq,Nagasaki:2012re}, providing a precision test of AdS/CFT in the presence of interface defects. Here we greatly extend the prior analysis to obtain the planar $\la \cO_J\ra_\cD$ with exact $\lambda$ dependence for both $k=0$ and $k>0$. As we will explain, our exact expressions bridge  known weak coupling results from integrability methods and strong coupling answers from consideration of probe branes in IIB string theory on $AdS_5\times S^5$.

\subsection{The D5-brane interface at $k=0$}

For simplicity of the presentation, below we will first consider the case $k=0$ and then generalize to $k>0$.  

\subsubsection{Inserting bulk BPS operators}
As explained in Section~\ref{sec:dYM}, from the defect-Yang-Mills description \cite{Wang:2020seq}, the defect one-point function of $\cO_J$ inserted at the north pole of $S^4$ is computed by   an insertion of the form $\sum_{i=1}^N  f_J(a_i)$ in the  matrix model \eqref{k=0mm}, where $f_J$ is a degree $J$ polynomial in $a_i$ that takes into account potential mixing between the operator $\cO_J$ and its lower dimensional cousins on $S^4$.
The eigenvalue integral that we analyze is then given by\footnote{One can arrive at the same matrix model by doing a different localization computation using the  supercharge of \cite{Pestun:2007rz} in the $\cN=2$ subalgebra $\mf{osp}(4|2)$ of the $\cN=4$ SYM, taking into account the interface (boundary) defect  \cite{Drukker:2010jp,Bullimore:2014nla,Gava:2016oep,LeFloch:2017lbt,Bawane:2017gjf} as well as local operator insertion at the north pole \cite{Gerchkovitz:2014gta,Gerchkovitz:2016gxx}. There are two lessons here. One is that the agreement between the two localization computations is a consequence of the underlying $\cN=4$ superconformal symmetry of the theory. The other is that the matrix model computation we do here has a direct generalization for theories with only $\cN=2$ supersymmetry. It would be interesting to explore this further as an extension of extremal correlators in $\cN=2$ SCFTs \cite{Baggio:2014ioa,Baggio:2014sna} that incorporates interface and boundary defects. }
\beq\label{eq:unnormalcorr}
\langle  \mathcal{O}_J\rangle_\cD\equiv {\la\cD \cO_J \ra_{\rm SYM}\over \la\cD   \ra_{\rm SYM}}=\frac{1}{\mathcal{Z}}\int \left(\prod_{i=1}^{N}da_i\right) \frac{\Delta (a)}{\prod_{k}2\cosh \pi a_i}\left[\sum_{i=1}^{N}f_J(a_i)\right] \,\,e^{-\frac{8\pi^2 N}{\lambda}\sum_{k}a_i^2}\,,
\eeq
where the expectation value of the interface defect $\la \cD\ra_{\rm SYM}$ is computed by taking the ratio of $\cZ$ and the partition function of SYM without the interface insertion.

As explained in \cite{DeWolfe:2001pq,Wang:2020seq}, the residual $\mf{so}(3)_{567}\times \mf{so}(3)_{890}$ R-symmetry of the half-BPS interface defect implies that the one-point function $\la \cO_J\ra_\cD$ is only nonzero for even $J$. This can also be seen readily from the above matrix model. 

A convenient choice of the function $f_J(a_i)$ is determined by diagonalizing the two-point function of the operators $\cO_J$ using the Gram-Schmidt procedure. As shown in the paper \cite{Rodriguez-Gomez:2016cem}, the result reads
\beq
\begin{aligned}
	f_J(a)=g^{J}\left[2T_{J} \left(\frac{a}{2g}\right)+\delta_{J,2}\right]\comma
\end{aligned}
\eeq
where $g$ is defined by\fn{This is the standard notation in the integrability literature, although it is confusing at times.}
\beq
g=\frac{\sqrt{\lambda}}{4\pi}\comma 
\eeq
and $T_n$ is the Chebyshev polynomial. Note that these single-trace operators are not canonically normalized. Namely their two-point functions are given by
\beq\label{eq:2ptnorm}
\langle \mathcal{O}_J\bar{\mathcal{O}}_{J}\rangle_{\rm SYM}=J(2g)^{2J}\period
\eeq
The canonically normalized operators $\mathcal{O}^{\rm norm}_J$ are defined by
\beq
\mathcal{O}^{\rm norm}_J\equiv \frac{i^{J}}{2^{\frac{J}{2}}g^{J}\sqrt{J}}\mathcal{O}_J\comma
\eeq
where the factor $i^{J}$ is purely a convention we chose to match the result with the results in perturbation theory. After the normalization the two-point function reads
\beq
\langle \mathcal{O}^{\rm norm}_J\bar{\mathcal{O}}_{J}^{\rm norm}\rangle_{\rm SYM} =2^{J}\period
\eeq
Here $2^{J}$ should be understood as coming from the R-symmetry polarization vectors $(Y\cdot \bar{Y})^{J}$.

To compute the correlator \eqref{eq:unnormalcorr}, it is simpler to exponentiate the piece describing the local operator and consider the following matrix model
\beq
\mathcal{Z}_J\equiv \int [da] \frac{\Delta (a)}{\prod_{i=1}^N2\cosh \pi a_i}\,\,\exp \left[-\frac{N}{2g^2}\sum_{i=1}^Na_i^2-Ng_{J}\sum_{i=1}^{N}f_J(a_i)\right]\period
\label{k=0mmf}
\eeq
The logarithmic derivative of this partition function gives the normalized ratio
\beq\label{eq:logderivative}
\la \cO_J\ra_\cD=-\frac{1}{N}\left.\del_{g_J}\log \mathcal{Z}_{J}\right|_{g_J=0} \period
\eeq

\subsubsection{Equations for resolvent}
We will solve the matrix model \eqref{k=0mmf} in the planar large $N$ limit using standard techniques \cite{Brezin:1977sv,Eynard:2015aea}. For this purpose, we introduce a normalized density of eigenvalues
\beq
\rho (a-a_i)\equiv \frac{1}{N}\sum_{i=1}^{N}\delta (a-a_i)\comma
\eeq
and express the integral as
\beq
\mathcal{Z}_J=\int \left(\prod_{i=1}^{N}\frac{da_i}{2\cosh\pi a_i}\right)\exp \left[-N^2S_{\rm eff}[\rho]\right] \comma
\eeq
with
\beq
S_{\rm eff}[\rho]=\frac{1}{2g^2}\int da \rho(a)a^2+g_J\int da \rho(a)f_J(a)-\frac{1}{2}\dashint da db\,\rho (a)\rho(b)\log (a-b)^2 \period
\eeq
Here the last term should be interpreted as the principal value integral. Taking the variation with respect to $\rho(a)$ and further taking $\del_a$, we get the saddle point equation
\beq\label{eq:saddle}
\frac{a}{g^2}+g_Jf_J^{\prime}(a)=2 \dashint db\frac{\rho (b)}{a-b}\period
\eeq
Here again, the integral is interpreted as the principal value integral. Namely we have
\beq
\dashint db\frac{\rho (b)}{a-b} =\frac{1}{2}\int db \,\rho (b)\left[\frac{1}{a-b+i\epsilon}+\frac{1}{a-b-i\epsilon}\right]\period
\eeq

To proceed, we introduce the resolvent
\beq
R (a)=\frac{1}{N}\sum_{i=1}^{N}\frac{1}{a-a_i}=\int db\frac{\rho(b)}{a-b}\comma
\eeq
which satisfies for $\ep \to 0^+ $
\beq\label{eq:rhoR}
-\frac{1}{2\pi i}\left[R(a+i\epsilon)-R (a-i\epsilon)\right]=\rho(a)\period
\eeq
Using \eqref{eq:rhoR}, we can rewrite \eqref{eq:saddle} as
\beq
R(a+i\epsilon)+R(a-i\epsilon)=\frac{a}{g^2}+g_Jf_J^{\prime}(a)\period
\eeq

In addition to this equation, we also need to impose the normalization condition $\int \rho(a)da=1$. This is equivalent to imposing that $R(a)$ decays as $1/a$ at infinity. Thus, to summarize, the equations we need to solve are
\begin{shaded}
	\begin{align}
	&R(a+i\epsilon)+R(a-i\epsilon)=\frac{a}{g^2}+g_Jf_J^{\prime}\label{eq:disconteq}\comma\\
	&R(a)\sim \frac{1}{a}\qquad a\to \infty\period\label{eq:asympteq}
	\end{align}
\end{shaded}
\subsubsection{Solving the matrix model}
Since we are interested in a small deformation of the Gaussian matrix model by $g_J$, we can assume that the eigenvalue in the large $N$ limit condenses into a single cut $[\mu_{-},\mu_{+}]$.

Now, consider the following integral
\beq\label{eq:contourR}
I=\oint_{\mathcal{C}} \frac{dv}{2\pi i}\sqrt{\frac{(u-\mu_{+})(u-\mu_{-})}{(v-\mu_{+})(v-\mu_{-})}}\frac{R(v)}{u-v}\comma
\eeq 
where the contour $\mathcal{C}$ encircles the branch cut $[\mu_{-},\mu_{+}]$ and $u$ is outside the contour. By deforming the contour and sending it to infinity, we pick up the contribution from a pole at $u=v$ and get
\beq
I=R(u)\period
\eeq 
Note that the contribution from infinity can be neglected owing to the asymptotic behavior of $R(u)$ \eqref{eq:asympteq}.
On the other hand, using \eqref{eq:disconteq}, we can evaluate $I$ alternatively as
\beq
\begin{aligned}
	I&=\int_{\mu_{-}-i\epsilon}^{\mu_{+}-i\epsilon}\frac{dv}{2\pi i} \sqrt{\frac{(u-\mu_{+})(u-\mu_{-})}{(v-\mu_{+})(v-\mu_{-})}}\frac{R(v+i\epsilon)+R(v-i\epsilon)}{u-v}\comma\\
	&=\int_{\mu_{-}-i\epsilon}^{\mu_{+}-i\epsilon}\frac{dv}{2\pi i} \sqrt{\frac{(u-\mu_{+})(u-\mu_{-})}{(v-\mu_{+})(v-\mu_{-})}}\frac{\frac{v}{g^2}+g_{J}f_J^{\prime}(v)}{u-v}\comma
\end{aligned}
\eeq
which can be converted back to the contour integral
\beq
I=\oint\frac{dv}{4\pi i} \sqrt{\frac{(u-\mu_{+})(u-\mu_{-})}{(v-\mu_{+})(v-\mu_{-})}}\frac{\frac{v}{g^2}+g_{J}f_J^{\prime}(v)}{u-v}\period
\eeq
Equating the two expressions, we obtain an integral representation of $R(u)$,
\beq\label{eq:integralrep}
R(u)=\oint\frac{dv}{4\pi i} \sqrt{\frac{(u-\mu_{+})(u-\mu_{-})}{(v-\mu_{+})(v-\mu_{-})}}\frac{\frac{v}{g^2}+g_{J}f_J^{\prime}(v)}{u-v}\period
\eeq
This is just a standard solution for the Riemann-Hilbert problem \eqref{eq:disconteq}.

\paragraph{Gaussian matrix model} Now the remaining task is to determine the position of the branch points $\mu_{\pm}$. Let us first review how this can be achieved for the Gaussian matrix model, namely for $g_{J}=0$ \cite{Brezin:1977sv,Eynard:2015aea}. In this case, owing to the parity symmetry of the potential, we can assume $\mu_{+}=-\mu_{-}=\mu$. We then have
\beq
R_{\rm Gauss}(u)=\frac{1}{g^2}\oint\frac{dv}{4\pi i} \sqrt{\frac{u^2-\mu^2}{v^2-\mu^2}}\frac{v}{u-v}\period
\eeq
The right hand side can be evaluated simply by deforming the contour to infinity. As a result we get
\beq
R(u)=\frac{1}{2g^2}\left(\underbrace{u}_{v=u}-\underbrace{\sqrt{u^2-\mu^2}}_{v=\infty}\right)\comma
\eeq
where, as indicated, the first term comes from the pole at $v=u$ while the second term comes from the pole at infinity. To determine $\mu$, we then impose the condition \eqref{eq:asympteq} on the asymptotic behavior $R(u)\sim1/u$. We thus find $\mu=2g$, and
\beq
R(u)=\frac{1}{2g^2}\left(u-\sqrt{u^2-4g^2}\right)=\frac{1}{g\, x(u)}\period
\eeq
Here $x(u)$ is the Zhukovsky variable, which is commonly used in the literature on integrability. Its definition is given by
\beq\label{eq:Zhukowskydef}
u=g\left(x(u)+\frac{1}{x(u)}\right)\quad \iff \quad x(u)=\frac{u+\sqrt{u^2-4g^2}}{2g}\period
\eeq

\paragraph{Infinitesimal deformation} We now consider infinitesimal deformation of the eigenvalue potential in the matrix model
\beq
V^{\prime}(u)=\frac{u}{g^2}\quad \to \quad \frac{u}{g^2}+ g_J f^{\prime}(u)\comma
\eeq
with $g_J\ll 1$. Differentiating the equation \eqref{eq:integralrep} with respect to $g_J$, we get
\begin{align}
	\frac{\del R(u)}{\del g_J}=&\oint\frac{dv}{4\pi i} \sqrt{\frac{(u-\mu_{+})(u-\mu_{-})}{(v-\mu_{+})(v-\mu_{-})}}\frac{f_J^{\prime}(v)}{u-v}\\
	&+\frac{1}{g^2}\oint\frac{vdv}{8\pi i} \sqrt{\frac{(u-\mu_{+})(u-\mu_{-})}{(v-\mu_{+})(v-\mu_{-})}}\left[\frac{\del_{g_J}\mu_{+}}{(v-\mu_{+})(u-\mu_{+})}+\frac{\del_{g_J}\mu_{-}}{(v-\mu_{-})(u-\mu_{-})}\right]\period\nn
\end{align}
Since we are interested in a small deformation away from the Gaussian point, we can set $\mu_{+}=-\mu_{-}=2g$ in the formula and get
\beq
\begin{aligned}
	\left.\frac{\del R(u)}{\del g_J}\right|_{g_J=0}=&\oint\frac{dv}{4\pi i} \sqrt{\frac{u^2-4g^2}{v^2-4g^2}}\frac{f_J^{\prime}(v)}{u-v}\\
	&+\frac{1}{g^2}\oint\frac{vdv}{8\pi i} \sqrt{\frac{u^2-4g^2}{v^2-4g^2}}\left[\frac{\del_{g_J}\mu_{+}}{(v-2g)(u-2g)}+\frac{\del_{g_J}\mu_{-}}{(v+2g)(u+2g)}\right]\period
\end{aligned}
\eeq
To compute these integrals, we express them using the Zhukovsky variables
\beq
x \equiv x(u)\comma\qquad y\equiv x(v)\period
\eeq
Note that $f_J(v)$ takes the following simple form in terms of $y$ as
	\beq
	\begin{aligned}
		f_J(v)&=g^{J}\left(2T_J(v/2g)+\delta_{J,2}\right)=g^{J}\left[2T_J\left(\frac{y+1/y}{2}\right)+\delta_{J,2}\right]= g^{J}\left[y^{J}+\frac{1}{y^{J}}+\delta_{J,2}\right]\comma
	\end{aligned}
	\eeq
	 where we have used the identity $T_J(\cos \theta)=\cos J \theta$, and $dv$ can be rewritten as
\beq\label{eq:conversion}
dv=dy\frac{dv}{dy}=dyg\left(1-\frac{1}{y^2}\right)\period
\eeq

As a result, we get
\beq
\begin{aligned}
	\left.\frac{\del R(u)}{\del g_J}\right|_{g_J=0}=\begin{cases}\frac{g^{J-1} J(1-x^{-J})}{x-x^{-1}}+\frac{1}{4g^2}\left[\frac{x+1}{x-1}\del_{g_J}\mu_{+}+\frac{x-1}{x+1}\del_{g_J}\mu_{-}\right]&\qquad J:{\rm even}\\\frac{g^{J-1} J(\frac{x+x^{-1}}{2}-x^{-J})}{x-x^{-1}}+\frac{1}{4g^2}\left[\frac{x+1}{x-1}\del_{g_J}\mu_{+}+\frac{x-1}{x+1}\del_{g_J}\mu_{-}\right]&\qquad J:{\rm odd}\end{cases}\period
\end{aligned}  
\eeq
To determine $\del_{g_{J}}\mu_{\pm}$, we impose the condition $R(u)\sim 1/u$. This translates to the following condition on $\del_{g_J} R (u)$:
\beq
\left.\frac{\del R (u)}{\del g_J}\right|_{g_J=0}\sim O(1/u^2) \qquad u\to \infty\period
\eeq
We then obtain
\beq
\begin{aligned}
	J:\text{even} \qquad &\del_{g_J}\mu_{+}=-\del_{g_J}\mu_{-}=-g^{J+1}J\comma\\
	J:\text{odd} \qquad &\del_{g_J}\mu_{+}=\del_{g_J}\mu_{-}=-g^{J+1}J\period
\end{aligned}
\eeq
Putting the above together, the derivative of the resolvent is given by
\begin{shaded}
	\beq
	\left.\frac{\del R(u)}{\del g_{J}}\right|_{g_{J}=0}=-\frac{g^{J-1}J}{x^{J}}\frac{1}{x-x^{-1}}\comma
	\eeq
\end{shaded}
\noindent  both for even and odd $J$.
\subsubsection{Computation of defect one-point functions}
Let us now compute the one-point function in the presence of the interface defect using the formula \eqref{eq:logderivative}. In the large $N$ limit, the modified partition function $\mathcal{Z}_J$ is given by 
\beq
\mathcal{Z}_{J} \sim e^{-N^{2}S_{\rm eff}[\rho]} e^{-N\int\frac{du}{2\pi i} R(u)\log (2\cosh \pi u) }\comma
\eeq
where $\rho(u)$ and $R (u)$ are saddle point values of the density and the resolvent respectively. By taking the logarithmic derivative, we get
\beq\label{eq:logder2}
-\frac{1}{N}\del_{g_J}\log \mathcal{Z}_J =N \frac{dS_{\rm eff}}{dg_J}+ \oint \frac{du}{2\pi i}\del_{g_J}R(u)\log (2\cosh \pi u)\period
\eeq
Among the two terms on the right hand side, the first term can be decomposed into two contributions, one coming from the explicit $g_J$ dependence of the action and the other coming from the variation of $\rho$, $\del_{g_J}\rho$:
\beq
\frac{d S_{\rm eff}}{dg_{J}}=\frac{\del S_{\rm eff}}{\del g_J} +\underbrace{\frac{\del \rho}{\del g_J} \frac{\delta S_{\rm eff}}{\delta \rho}}_{=\,0}\period
\eeq
As indicated, the second contribution vanishes owing to the equation of motion. It turns out that the first term also vanishes precisely because of our choice of $f_J$:
\beq
\begin{aligned}
	\left.\frac{\del S_{\rm eff}}{\del g_J} \right|_{g_J=0}&=-\oint \frac{du}{2\pi i}R(u) f_{J}(u)\,,\\
	&=-\oint \frac{dx}{2\pi i}g \left(1-\frac{1}{x^2}\right)\frac{1}{g x}g^{J}\left(x^{J}+\frac{1}{x^{J}}+\delta_{J,2}\right)=0\period
\end{aligned}
\eeq
In the second line, we expressed the integral in terms of the Zhukovsky variable using the conversion formula \eqref{eq:conversion} and the integration contour is along the unit circle.
Therefore, we only need to compute the second term in \eqref{eq:logder2}. The result reads
\beq\label{eq:ratio}
\begin{aligned}
	\la \cO_J \ra_\cD &=-g^{J-1}J\oint \frac{du}{2\pi i}\frac{1}{x^{J}}\frac{\log (2\cosh \pi u)}{x-x^{-1}}\,,\\
	&=-g^{J}J\oint \frac{dx}{2\pi i x^{J+1}}\log \left[2\cosh \pi  g \left(x+\frac{1}{x}\right)\right]\period
\end{aligned}
\eeq
The residue is clearly zero for $J$ odd thus $\la \cO_J \ra_\cD$ is nonzero only for $J$ even as explained in the beginning of the section.

\paragraph{Final result}
The structure constant of the defect CFT can be computed from \eqref{eq:ratio} by dividing by the square root of the two-point function of the operator $\mathcal{O}_J$ \eqref{eq:2ptnorm}. The result reads\footnote{One can in principle expand the logarithm in \eqref{eq:final} and evaluate the integral term by term. This would lead to an infinite sum representation of \eqref{eq:final}. However we did not find it particularly useful for the purpose of this paper.}
\begin{shaded}
	\beq\label{eq:final}
	c_J\equiv \la \mathcal{O}^{\rm norm}_J \ra_\cD=-\frac{i^{J}}{2^{J/2}}\sqrt{J}\oint \frac{dx}{2\pi i x^{J+1}}\log \left[2\cosh \pi  g \left(x+\frac{1}{x}\right)\right]\period
	\eeq
\end{shaded}
\noindent An alternative expression can be obtained by performing integration by parts and changing the integration variable $x\to 1/x$:
	\beq
	\begin{aligned}
		c_J&=\frac{i^{J}}{2^{J/2}\sqrt{J}}\oint \frac{dx}{2\pi i }g\left(1-\frac{1}{x^2}\right) \pi x^{J}\tanh\left[ \pi  g \left(x+\frac{1}{x}\right)\right]\,.
	\end{aligned}
	\eeq

We can also compute the expectation value of the interface defect at large $N$. The result reads
\begin{shaded}
\beq
\log \langle \mathcal{D}\rangle_{\rm SYM}=-\frac{N}{g}\oint \frac{dx}{2\pi i x}\log \left[2\cosh \pi g \left(x+\frac{1}{x}\right)\right]\period
\eeq
\end{shaded}

\subsubsection{Weak coupling expansion}
The expansion at weak coupling can be obtained straightforwardly by expanding the integral \eqref{eq:final}. The result  for even $J$ reads
\beq
c_J=-\left(ig\pi\right)^{J}2^{\frac{J}{2}}\frac{B_J(2^{J}-1)}{\Gamma (J+1)\sqrt{J}}+\cdots\comma
\eeq
where $B_J$ is the Bernoulli number.
\subsubsection{Strong coupling expansion}
To compute the result at strong coupling, we rewrite the integral (for even $J$) as
\beq
\begin{aligned}
	c_J&=-2\frac{i^{J}}{2^{J/2}}\sqrt{J}\int_{-\pi/2}^{\pi/2} \frac{d\theta}{2\pi }e^{-iJ \theta}\log \left[2\cosh (2\pi g \cos\theta)\right]\,,\\
	&=-2\frac{i^{J}}{2^{J/2}}\sqrt{J}\int_{-\pi/2}^{\pi/2} \frac{d\theta}{2\pi }e^{-iJ \theta}\left(2\pi g \cos \theta+\log \left[1+e^{-4\pi g \cos\theta}\right]\right)\period
\end{aligned}
\eeq
Since the second term in the parenthesis is exponentially suppressed\fn{Note that $\cos \theta>0$ in the integration range of the integral.}, we simply need to evaluate the first term,
\beq\label{eq:strongk=0}
\begin{aligned}
	c_J&\sim -4\frac{i^{J}}{2^{J/2}}\pi g\sqrt{J}\int_{-\frac{\pi}{2}}^{\frac{\pi}{2}}\frac{d\theta}{2\pi} e^{-i J\theta} \cos \theta=\frac{4 g(-1)^{J}\sqrt{J}}{2^{\frac{J}{2}}(J^2-1)}\period
\end{aligned}
\eeq
This is in agreement with the finding in \cite{Wang:2020seq}.

\subsection{Generalization to D5-brane interface at $k>0$}
The matrix model computation for the $k> 0$ interface is a straightforward generalization of what we have done in the previous section. 
We start with the relevant matrix integrals for $k>0$ from \eqref{D5mmsim} with the insertion $\cO_J$ on the north pole, dropping (common) unimportant overall coefficients,
\begin{shaded}
	\begin{align}
	\langle \mathcal{D}\mathcal{O}_J\rangle_{k}=&\int [da]\frac{\Delta (a)^2}{\prod_{j=1}^N 2\cosh \pi \left(a_j\red{+ {k i\over 2}} \right)}\left[\sum_{j=1}^{N}f_J(a_j)+\red{\sum_{s=-\frac{k-1}{2}}^{\frac{k-1}{2}}f_{J}(i s)}\right]\nonumber\\
	&\times \red{\prod_{s=-\frac{k-1}{2}}^{\frac{k-1}{2}}\prod_{n}\left(a_j-is\right)} e^{-\frac{8\pi^2 N}{\lambda }\sum_{i=1}^N a_i^2}\comma\\
	\langle \mathcal{D}\rangle_{k}=&\int [da] \frac{\Delta (a)^2}{\prod_{j=1}^N 2\cosh \pi \left(a_j\red{+ {k i\over 2}} \right)}\red{\prod_{s=-\frac{k-1}{2}}^{\frac{k-1}{2}}\prod_{j=1}^N\left(a_j-is \right)} e^{-\frac{8\pi^2 N}{\lambda}\sum_{i=1}^N a_i^2}\period
	\end{align}
	\end{shaded}
\noindent
Here we have denoted in red the modifications of the matrix model compared to the $k=0$ case. Note that if the local operator $\cO_J$ is inserted at the south pole, the only difference would be dropping the red term in the square bracket above. 

In the presence of these extra factors, the logarithmic derivative of the $k>0$ interface partition function receives the following extra contributions
\beq
\la \cO \ra_{\cD_k}\equiv \frac{\langle \mathcal{D}\mathcal{O}_J\rangle_k}{\langle \mathcal{D}\rangle_k}= \left.\la \cO_J \ra_\cD \right|_{\rm modified} \underbrace{-\sum_{s=-\frac{k-1}{2}}^{\frac{k-1}{2}}\oint \frac{du}{2\pi i} \del_{g_J}R(u)\log \left(u-is\right)}_{\equiv \,{\tt extra1}}\underbrace{+\sum_{s=-\frac{k-1}{2}}^{\frac{k-1}{2}}f_J(is)}_{\equiv \,{\tt extra2}}\period
\eeq
Here the first term on the RHS comes from modifying the result for $k=0$ by taking into account the shift in $\cosh \pi  a_j  \to \cosh \left(\pi a_j+ {k i\over 2}\pi \right) $ in the matrix model \eqref{D5mmsim},
\ie
\left.\la \cO_J \ra_\cD \right|_{\rm modified}=
\oint \frac{dx}{2\pi i}g\left(1-\frac{1}{x^{2}}\right) \,x^{J}\,\pi \tanh \left[\pi g \left(x+\frac{1}{x}\right)+{ k i\over 2}\pi\right]\,.
\fe
Using the explicit form of $\del_{g_J}R$, the first extra contribution can be evaluated as
\beq
\begin{aligned}
	{\tt extra1}&=g^{J}J\sum_{s=-\frac{k-1}{2}}^{\frac{k-1}{2}}\oint \frac{dx}{2\pi i}\left(1-\frac{1}{x^2}\right)\frac{1}{x^{J}}\frac{\log\left[g((x+x^{-1})-(x_s+x_s^{-1}))\right]}{x-x^{-1}}\,,\\
	&= g^J\sum_{s=-\frac{k-1}{2}}^{\frac{k-1}{2}}\oint_{|x|=1}\frac{dx}{2\pi i x^{J}}\left(1-\frac{1}{x^2}\right)\frac{1}{x+x^{-1}-(x_s+x_s^{-1})}\,,
\end{aligned}
\eeq
where $x_s \equiv  x(is)$. We can evaluate this integral by pushing the contour to infinity, and picking the residues at poles at $x=x_s$. Then we get
\beq
{\tt extra1}= -g^{J}\sum_{s=-\frac{k-1}{2}}^{\frac{k-1}{2}} x_s^{-J}\period
\eeq
On the other hand, the second extra contribution can be expressed as
\beq
\begin{aligned}
	{\tt extra2}&=g^{J}\sum_{s=-\frac{k-1}{2}}^{\frac{k-1}{2}}\left[x_s^{J}+x_s^{-J}+\delta_{J,2}\right]\,.
\end{aligned}
\eeq
We thus conclude that the sum of extra contributions is given by
\beq
{\tt extra1}+{\tt extra2}=g^{J}\sum_{s=-\frac{k-1}{2}}^{\frac{k-1}{2}}\left(x_s^{J}+\delta_{J,2}\right)\period
\eeq

Taking into account the normalization of the operator, we thus get the following result at large $N$:
\begin{shaded}
	\beq
	\begin{aligned}
		c_J^{(k)}&=\frac{i^{J}}{2^{\frac{J}{2}}\sqrt{J}}\left({\tt integral}+{\tt sum}\right)\comma\label{eq:finalck}
	\end{aligned}
	\eeq
	with
	\begin{align}
	{\tt integral}&=\begin{cases}
	\oint \frac{dx}{2\pi i}g\left(1-\frac{1}{x^{2}}\right) \,x^{J}\,\pi \coth \left[\pi g \left(x+\frac{1}{x}\right)\right] & k \in 2\mZ+1\comma 
	\\
	\oint \frac{dx}{2\pi i}g\left(1-\frac{1}{x^{2}}\right) \,x^{J}\,\pi \tanh \left[\pi g \left(x+\frac{1}{x}\right)\right] & k \in 2\mZ\comma 
	\end{cases}\\
	{\tt sum}&=\sum_{s=-\frac{k-1}{2}}^{\frac{k-1}{2}}\left(x_s^{J}+\delta_{J,2}\right)\period
	\end{align}
\end{shaded}

\subsection{Comparisons with results from perturbation theory and holography}
\subsubsection{Weak coupling expansions }
Let us first expand the result \eqref{eq:finalck} at weak coupling. The contribution from the integral part ${\tt integral}$     to $c_J^{(k)}$ starts at $\cO(g^{J})$, while the contribution from the ${\tt sum}$ starts at $\cO(g^{-J})$ which follows from the expansion
\beq
x(u)=\frac{u-\frac{g^2}{u}+\cdots}{g}\period
\eeq
Let us now expand $c_J^{(k)}$ up to the subleading order $\cO(g^{2-J})$ in order to compare with the perturbative results in \cite{Buhl-Mortensen:2016jqo} using Feynman diagrams at tree level and one-loop. For this purpose we can just focus on the contribution from the ${\tt sum}$ part. 

Using the following summation identity for the Bernoulli Polynomial $B_n(z)$
\ie
\sum_{s=-{k-1\over 2}}^{ {k-1\over 2}} s^J=-\frac{2}{J+1} B_{J+1}\left({1-k\over 2}\right)\,,
\fe
we obtain  
	\beq
	c_J^{(k)}=-\frac{2}{2^{\frac{J}{2}}g^J\sqrt{J}}\left[\frac{B_{J+1}\left(\frac{1-k}{2}\right)}{J+1}+g^2\frac{JB_{J-1}\left(\frac{1-k}{2}\right)}{J-1}+g^2\delta_{J,2} {k\over 2} \right]\period
	\eeq
This beautifully matches the result in \cite{Buhl-Mortensen:2016jqo}.\footnote{Note that the results there are not properly normalized.}
\subsubsection{Strong coupling expansions and holography}
Here we consider the strong coupling limit of the defect one-point function $\la \cO_J \ra_{\cD_k}$ where we draw connection to computations in IIB string theory on $AdS_5\times S^5$ via the holographic correspondence \cite{Maldacena:1997re}. After taking the near horizon limit of the D3-branes, the D5-brane interface defect that interpolates between the $SU(N)$ and $SU(N+k)$ SYMs maps to a probe D5-brane wrapping an $AdS_4\times S^2$ submanifold in the bulk with $k$-units of worldvolume flux through the $S^2$ factor \cite{Karch:2000gx}.\footnote{In the leading large $N$ limit, the interface between the $SU(N)$ and $SU(N+k)$ SYMs is indistinguishable from that between the $U(N)$ and $U(N+k)$ SYMs which we have studied in previous sections.} The embedding of the submanifold  $AdS_4\times S^2$ in $AdS_5\times S^5$ depends on the spacetime and R-symmetry orientation of the interface defect on the boundary (here we follow the convention in \cite{Wang:2020seq}). The embedding $AdS_4 \subset AdS_5$ also depends on $k$ \cite{Karch:2000gx}. The one-point function  $\la \cO_J \ra_{\cD_k}$ can be computed by standard holographic methods in the bulk which we review below and match with limits of our exact expression \eqref{eq:finalck} in the 't Hooft coupling $\lambda$ and flux quanta $k$.

\paragraph{Large $J$} The strong coupling behavior of the integral part ${\tt integral}$ was already evaluated in \eqref{eq:strongk=0}. Therefore, we just need to know the strong-coupling behavior of the sum part ${\tt sum}$. In the scaling limit $k\sim \sqrt{\lambda}$, we can approximate the sum by an integral
\beq
\begin{aligned}\label{eq:integraltoev}
	{\tt sum}&\sim 2g \int_{0}^{2\kappa} dz x(ig z)^{J}
\end{aligned}
\eeq
where $\kappa$ is the ratio
\beq
\kappa =\frac{\pi k}{\sqrt{\lambda}}=\frac{k}{4g}\period
\eeq
Since each term in the integrand is monotonic in $z$, we can evaluate them at the edges of the integration range when $J\gg 1$. As a result, we get
\beq
{\tt sum}\sim 2gi^{J}\left(\kappa+\sqrt{\kappa^2+1}\right)^{J}\period 
\eeq
Since  $J\gg 1$, this contribution is exponentially larger than the contribution from the integral ${\tt integral}$ given in \eqref{eq:strongk=0}.

Multiplying the prefactor in \eqref{eq:finalck}, we see that it reproduces the result computed from the classical string worldsheet \cite{Buhl-Mortensen:2015gfd}
\beq
c_J^{(k)}\sim \left(\frac{\kappa+\sqrt{\kappa^2+1}}{\sqrt{2}}\right)^{J}\period
\eeq

\paragraph{Finite $J$}
For finite $J$, we can simply perform the integral \eqref{eq:integraltoev} analytically to get
\beq
{\tt sum}=-\frac{4g J}{J^2-1}+(\kappa+\sqrt{1+\kappa^2})^{J}\frac{4g\left[-\kappa+J\sqrt{1+\kappa^2}\right]}{J^2-1}\period
\eeq
It turns out that the first term cancels the integral term ${\tt integral}$ given in \eqref{eq:strongk=0} and the final answer reads
	\beq
	c_J^{(k)} =\left(\kappa+\sqrt{1+\kappa^2}\right)^{J}\frac{4(-1)^{J}g \left[-\kappa+J\sqrt{1+\kappa^2}\right]}{2^{\frac{J}{2}}\sqrt{J}(J^2-1)}\,.
	\label{cJstrong}
	\eeq

The bulk Witten diagram computation of the one-point function $c_J^{(k)}$ was carried out at tree-level in IIB supergravity in \cite{Nagasaki:2011ue}  by taking into account interaction vertices on the world-volume of the probe D5-brane. The result takes the form (see also \cite{Wang:2020seq})
\beq
c_J=\mathcal{C}_{\frac{J}{2}}g\frac{  2^{2+\frac{J}{2}}\Gamma \left(J+\frac{1}{2}\right)}{\sqrt{\pi J}\Gamma (J)}\int_0^{\infty}du \frac{u^{J-2}}{\left[(1-\kappa u)^2+u^2\right]^{J+\frac{1}{2}}}\comma
\eeq
with $\mathcal{C}_{\frac{J}{2}}$  coming from integrating the internal part of the bulk wavefunction over the $S^2$  
\beq
\mathcal{C}_{\frac{J}{2}}={1\over 4\pi}\int dV_{S^2} (\cos \theta)^{J\over 2}=   \frac{(-1)^{J}}{J+1}\period
\eeq
The integral can be performed for each integer $J$ analytically. By computing it for various different values and using ${\tt FindSequenceFunction}$ in Mathematica, we found that the result is given by
\beq
\begin{aligned}
	&\int_0^{\infty}du \frac{u^{J-2}}{\left[(1-\kappa u)^2+u^2\right]^{J+\frac{1}{2}}}=\\
	&\left(\kappa+\sqrt{1+\kappa^2}\right)^{J-1}\frac{\sqrt{\pi}\Gamma (J-1)}{2^{J+1}\Gamma\left(J+\frac{1}{2}\right)}\left[J+1+(J-1)(\kappa+\sqrt{1+\kappa^2})^2\right]\period
\end{aligned}
\eeq
Combining all the factors, we get precisely \eqref{cJstrong}.

\section{Integrable Bootstrap for One-Point Functions}
\label{sec:integrable}
In this section, we focus on the planar limit and study the defect one-point functions from integrability. 
The defect one-point functions of non-BPS single-trace operators were studied at weak coupling in \cite{deLeeuw:2015hxa,Buhl-Mortensen:2015gfd,Buhl-Mortensen:2016pxs,deLeeuw:2016umh,Buhl-Mortensen:2016jqo,deLeeuw:2016ofj,Buhl-Mortensen:2017ind,deLeeuw:2017cop,deLeeuw:2018mkd,Grau:2018keb,Gimenez-Grau:2019fld,deLeeuw:2019ebw,Widen:2018nnu}.  The results for the D5-brane defect exhibit two important~features: 
\begin{enumerate}
\item They are nonzero only when the single-trace operator corresponds to a parity-symmetric Bethe state, $|u_1,-u_1,\ldots, u_{\frac{M}{2}},-u_{\frac{M}{2}}\rangle$, with $u$'s being the rapidities of the excitations on the spin chain. \item The results are given by a ratio of two determinants, each of which resembles the so-called Gaudin determinant \cite{tsuchiya1998determinant,brockmann2014gaudin,pozsgay2014overlaps,kozlowski2012surface}.
\end{enumerate} 
As explained in section 6 of \cite{Jiang:2019xdz}, these two features imply that the defect one-point functions can be interpreted as overlaps between an integrable boundary state on the string worldsheet and a closed string state describing the single-trace operator.

In what follows, we build on this assumption and bootstrap the integrable boundary state at finite $\lambda$ by imposing a set of consistency conditions.
\subsection{General strategy\label{subsec:generalstrategy}}
On the string-theory side, the one-point function in the presence of the D5-brane defect corresponds to a disk worldsheet with a closed-string vertex operator insertion. Viewed differently, it is an overlap between a closed string state $|\Psi\rangle$, which describes a single-trace operator, and the boundary state $|\mathcal{D}\rangle$, which describes the probe D5-brane in AdS:
\beq\label{eq:defdefect}
\langle \mathcal{O}\rangle_{\mathcal{D}}=\langle \mathcal{D}|\Psi \rangle\period
\eeq
In flat space such overlaps can be computed using the standard 2d CFT techniques. This is not the case in $AdS_5\times S^{5}$ since the worldsheet theory is strongly-coupled at finite $\lambda$. A way to overcome this problem is to use integrability. To apply integrability, we first gauge-fix the worldsheet diffeomorphism by choosing the generalized lightcone gauge.\footnote{See \cite{Arutyunov:2009ga} for a pedagogical review on the generalized lightcone gauge.} In this gauge, the spatial length of the string is proportional to one of the $R$-charges of the state (which we denote by $J$) and the resulting worldsheet theory is an integrable 2d theory of 8 massive bosons and 8 massive fermions \cite{Klose:2006zd,Arutyunov:2006yd}.

Assuming the boundary state $|\mathcal{D}\rangle$ is an integrable boundary state,\footnote{The integrable boundary states are defined as the boundary states which are annihilated by (infinitely many) odd-spin conserved charges \cite{Ghoshal:1993tm}. See also \cite{piroli2017integrable} for the analysis in integrable spin chains. } one can compute the overlap \eqref{eq:defdefect} following the strategy laid out in \cite{Jiang:2019xdz,Jiang:2019zig}:
\begin{enumerate}
\item First we consider the overlap for an infinitely long string $\langle \mathcal{D} |\Psi\rangle|_{J\to \infty}$. In this limit, the closed string state $|\Psi\rangle$ can be described as a collection of ``magnon'' excitations on the vacuum, for instance, as
\beq
|\Psi \rangle = |\mathcal{X}_1(u_1)\mathcal{X}_2(u_2)\cdots \mathcal{X}_M(u_M)\rangle
\eeq
where $\mathcal{X}_i$'s are magnons and $u_i$'s are their rapidities. Thus the right hand side of \eqref{eq:defdefect} is a function of these rapidities in this limit.

When $|\mathcal{D}\rangle$ is an integrable boundary state, the overlap in the infinite volume limit can be decomposed into two-particle overlaps
\beq
\langle \mathcal{D}|\mathcal{X}_1 (u)\mathcal{X}_2 (\bar{u}) \rangle
\eeq
where $\bar{u}$ is a parity-conjugate rapidity of $u$ (see \eqref{eq:defparity} for the definition).
\item Next we bootstrap the two-particle overlap by imposing a set of consistency conditions such as the global symmetry constraints, the Watson's equation, boundary Yang-Baxter equations, and crossing equations. Once determined, the two-particle overlaps allow us to write down the {\it asymptotic overlap}, which includes all the perturbative $1/J$ corrections.
\item To compute the overlaps at finite $J$ which include  corrections nonperturbative in $J$ (called the {\it wrapping corrections}), we first analyze the ground-state overlap $\langle \mathcal{D}|\Omega \rangle$ in the open string channel, and write down the thermodynamic Bethe ansatz (TBA). A crucial input for writing down the TBA is the reflection matrix, which can be obtained from the overlap determined in step 2 by the analytic continuation. From the TBA, we can derive the Fredholm-determinant representation \cite{Dorey:2004xk,Pozsgay:2010tv,Kostov:2018dmi,Kostov:2019sgu} for the ground-state overlap at finite $J$.
\item Finally, we generalize the result to excited states using the analytic continuation trick proposed by Dorey and Tateo \cite{Dorey:1996re}.
\end{enumerate}
In this paper, we perform the analysis up to step 2. This is enough for the comparison with weak-coupling results in the literature since the wrapping corrections kick in only at higher loop orders. In principle it should be possible to complete the program based on the results in this paper but we leave it to a future work. 

An important new ingredient which was not present in the analysis of \cite{Jiang:2019xdz} is the existence of {\it excited boundary states} (cf.~\cite{Ghoshal:1993tm}): The D5-brane defect does not correspond to a single boundary state, but rather corresponds to a set of boundary states, which can be viewed as excited states of some basic boundary state. In Section \ref{subsec:excitedB}, we show that this feature is essential in order to reproduce the results in the literature.  
\subsection{Dynamic $\mathfrak{psu}(2|2)$ spin chain}
Before delving into the actual computation, let us review the integrability description of closed-string states in $AdS_5\times S^{5}$. In the infinite volume limit $J\to \infty$, the closed string state is described by the dynamic $\mathfrak{psu}(2|2)$ spin chain introduced by Beisert \cite{Beisert:2005tm,Beisert:2006qh}. In this description, each magnon belongs to a bifundamental representation of the centrally-extended $\mathfrak{psu}(2|2)^2$ symmetry. The precise relation between the fields in $\mathcal{N}=4$ SYM and the excitations in the dynamic spin chain is given by
\beq
\begin{aligned}
&\phi^{a}\dot{\phi}^{\dot{b}}\mapsto \Phi^{a\dot{b}} &&(\Delta^{0}, J)=(1,0)\comma
&&\psi^{\alpha}\dot{\psi}^{\dot{\beta}}\mapsto D^{\alpha\dot{\beta}}Z&&(\Delta^{0}, J)=(2,1)\comma\\
&\psi^{\alpha}\dot{\phi}^{\dot{a}}\mapsto  \Psi^{\alpha\dot{a}}&&(\Delta^{0}, J)=(3/2,1/2)\comma
\qquad&&\phi^{a}\dot{\psi}^{\dot{\alpha}}\mapsto \Psi^{a\dot{\alpha}}&&(\Delta^{0}, J)=(3/2,1/2)\comma
\end{aligned}
\eeq
where the dotted and undotted indices correspond to the right and the left $\mathfrak{psu}(2|2)$ respectively and all the indices take $1$ or $2$. Here $\Psi$'s are fermion fields, $D^{\alpha\dot{\beta}}$'s are covariant derivatives, $Z$ and $\bar{Z}$ are the complex combination of the two scalars $\Phi_7+i \Phi_8$ and its conjugate, and $\Phi^{a\dot{b}}$'s are combinations of the other four scalars $\Phi_{5,6,9,0}$.  $\Delta^{0}$ is the classical dimension and $J$ is the $U(1)$ $R$-charge generated by $R_{78}$ in the full $SO(6)_R$ symmetry which we used to define the length of the string.\fn{$Z$ and $\bar{Z}$ have $+1$ and $-1$ charges for this $U(1)$ symmetry.} 
\paragraph{Symmetry}
Let us now summarize the action of the $\mathfrak{psu}(2|2)^2$ generators on the excitations. Since the actions of the left and the right $\mathfrak{psu}(2|2)$'s are identical, we only write the results for the left $\mathfrak{psu}(2|2)$:\fn{In what follows, $\epsilon^{12}=\epsilon^{\dot{1}\dot{2}}=-\epsilon_{12}=-\epsilon_{\dot{1}\dot{2}}=1$.}
\beq
\begin{aligned}
&R^{a}_{b} \ket{\phi^{c}}= \delta_{b}^{c}\ket{\phi^{a}}-\frac{1}{2}\delta^{a}_{b}\ket{\phi^{c}}\comma\quad
&&L^{\alpha}_{\beta}\ket{\psi^{\gamma}}=\delta^{\gamma}_{\beta} \ket{\psi^{\alpha}}-\frac{1}{2}\delta^{\alpha}_{\beta}\ket{\psi^{\gamma}}\comma\\
&Q^{\alpha}_{a}\ket{\phi^{b}}={\sf a}  \delta^{b}_{a}\ket{\psi^{\alpha}}\comma\quad &&Q^{\alpha}_{a}\ket{\psi^{\beta}}= {\sf b} \epsilon^{\alpha\beta}\epsilon_{ab}\ket{Z^{+}\phi^{b}}\comma\\
&S^{a}_{\alpha}\ket{\phi^{b}}= {\sf c} \epsilon^{ab}\epsilon_{\alpha\beta}\ket{Z^{-}\psi^{\beta}}\comma\quad &&S^{a}_{\alpha}\ket{\psi^{\beta}}={\sf d} \delta^{\beta}_{\alpha}\ket{\phi^{a}}\period 
\end{aligned}
\eeq
Here $L$'s are the Lorentz generators and $R$'s are the R-symmetry generators while $Q$'s and $S$'s are the supersymmetry and the superconformal generators respectively. The parameters ${\sf a}$-${\sf d}$ are functions of the rapidity $u$ of the excitation and are given by
\beq
{\sf a}(u)\equiv\sqrt{g}\gamma\comma \quad {\sf b}(u)\equiv \frac{\sqrt{g}}{\gamma}\left( 1-\frac{x^{+}}{x^{-}}\right)\comma \quad {\sf c}(u)\equiv\frac{i\sqrt{g} \gamma}{ x^{+}}\comma \quad {\sf d}(u) \equiv\frac{ \sqrt{g}x^{+}}{i\gamma}\left( 1- \frac{x^{-}}{x^{+}}\right)\comma\nonumber
\eeq
where $\gamma$ satisfies
\beq
|\gamma|^2 =  i(x^{-}-x^{+}) \period
\eeq
and $x(u)$ is the Zhukovsky variable \eqref{eq:Zhukowskydef} given by $u=g (x+1/x)$.
The plus and minus superscripts denote the shift of the rapidity by $i/2$, namely $f^{\pm}(u)=f(u\pm i/2)$.

In addition to these global charges, the dynamic spin chain has three central charges $C$, $P$ and $K$ which appear in the anti-commutators of fermionic charges
\beq
\begin{aligned}
&\{Q^{\alpha}{}_{a},Q^{\beta}{}_{b}\}=\epsilon^{\alpha\beta}\epsilon_{ab}P\comma&&\{\dot{Q}^{\dot{\alpha}}{}_{\dot{a}},\dot{Q}^{\dot{\beta}}{}_{\dot{b}}\}=\epsilon^{\dot{\alpha}\dot{\beta}}\epsilon_{\dot{a}\dot{b}}P\comma\\
&\{S^{a}{}_{\alpha},S^{b}{}_{\beta}\}=\epsilon^{ab}\epsilon_{\alpha\beta}K\comma
&&\{\dot{S}^{\dot{a}}{}_{\dot{\alpha}},\dot{S}^{\dot{b}}{}_{\dot{\beta}}\}=\epsilon^{\dot{a}\dot{b}}\epsilon_{\dot{\alpha}\dot{\beta}}K\comma\\
&\{Q^{\alpha}{}_{a},S^{b}{}_{\beta}\}=\delta^{b}_{a}L^{\alpha}{}_{\beta}+\delta^{\alpha}_{\beta}R^{b}{}_{a}+\frac{1}{2}\delta^{b}_{a}\delta^{\alpha}_{\beta}C\comma&&\{\dot{Q}^{\dot{\alpha}}{}_{\dot{a}},\dot{S}^{\dot{b}}{}_{\dot{\beta}}\}=\delta^{\dot{b}}_{\dot{a}}\dot{L}^{\dot{\alpha}}{}_{\dot{\beta}}+\delta^{\dot{\alpha}}_{\dot{\beta}}\dot{R}^{\dot{b}}{}_{\dot{a}}+\frac{1}{2}\delta^{\dot{b}}_{\dot{a}}\delta^{\dot{\alpha}}_{\dot{\beta}}C\period
\end{aligned}
\eeq
The action of these charges on the excitation $|\mathcal{X}\rangle$ reads 
\beq\label{eq:centralchargeaction}
\begin{aligned}
&C\ket{\mathcal{X}}= \frac{1}{2}({\sf a}{\sf d}+{\sf b}{\sf c})\ket{\mathcal{X}}\comma\quad 
P\ket{\mathcal{X}}={\sf a}{\sf b}\ket{Z\mathcal{X}}\comma \quad K\ket{\mathcal{X}}={\sf c}{\sf d}\ket{Z^{-}\mathcal{X}}\period
\end{aligned}
\eeq
Physically, $C$ is a linear combination of the dilatation $D$ and the $R$-symmetry charge $J$, 
\beq
C=\frac{D-J}{2}\comma
\eeq
while $P$ and $K$ correspond to the field-dependent gauge transformations (see e.g.~section 3.2 of \cite{Komatsu:2017buu} for further explanation). The extra insertions of $Z$ and $Z^{-1}$ are called $Z$-markers and are book-keeping devices for the nontrivial coproduct structure of the symmetry algebra \cite{Beisert:2005tm,Beisert:2006qh,Klose:2006zd,Arutyunov:2006ak}. On the field theory side they simply correspond to an insertion or a removal of the $Z$-field which can be moved around using the following rule: 
\beq
\begin{aligned}
&\ket{\mathcal{X}Z^{\pm}}=\left( \frac{x^{+}}{x^{-}}\right)^{\pm1}\ket{Z^{\pm}\mathcal{X}}\period
\end{aligned}
\eeq
\paragraph{Crossing, mirror and parity transformations} The energy and  the momentum of the magnon excitation admit compact expressions in terms of the Zhukovsky variable:
\beq\label{eq:energymomentum}
E(u)=\frac{1}{2}\frac{1+\frac{1}{x^{+}x^{-}}}{1-\frac{1}{x^{+}x^{-}}}\comma\qquad p(u)=\frac{1}{i}\log \frac{x^{+}}{x^{-}}\period
\eeq
Owing to the definition of the Zhukovsky variable \eqref{eq:Zhukowskydef}, they contain two branch cuts, one for $x^{+}$ and the other for $x^{-}$, when viewed as a function of the rapidity $u$. The analytic continuations across these branch cuts invert the corresponding Zhukovsky variables ($x^{\pm}\to 1/x^{\pm}$), and define analogues of the crossing and the mirror transformations in the relativistic field theory.

Let us first consider the analytic continuation (to be denoted by $u^{2\gamma}$) in which we cross both of the cuts once. This process transforms the Zhukovsky variables as
\beq
x^{+}(u^{2\gamma})=1/x^{+}(u)\comma\qquad x^{-}(u^{2\gamma})=1/x^{-}(u)\period
\eeq  
Using \eqref{eq:energymomentum}, one can check that this flips the signs of the energy and the momentum. Physically, this can be interpreted as the {\it crossing transformation}, which maps a particle to an antiparticle. 

If we instead cross only one of the two cuts, we have either
\beq
x^{+}(u^{\gamma})=1/x^{+}(u)\comma\qquad x^{-}(u^{\gamma})=x^{-}(u)\comma
\eeq
or 
\beq
x^{-}(u^{-\gamma})=x^{-}(u) \comma\qquad x^{-}(u^{-\gamma})=1/x^{-}(u)\comma
\eeq
depending on which cut we crossed. These transformations are interpreted as the {\it mirror transformations}. They map a particle in the original theory to a particle in the so-called mirror theory in which the roles of space and time on the worldsheet are swapped.

Yet another important transformation is the parity transformation $u\to \bar{u}$. This is nothing but the standard parity transformation on the worldsheet. In terms of the Zhukovsky variables, it is defined by
\beq\label{eq:defparity}
x^{+}(\bar{u})=-x^{-}(u)\comma\qquad x^{-}(\bar{u})=-x^{+}(u)\period
\eeq
One can readily check from \eqref{eq:energymomentum} that this flips the sign of the momentum but not of the energy.
 
\subsection{Symmetry constraints on the two-particle overlap}

\begin{figure}[t]
\centering
\begin{minipage}{0.45\hsize}
\centering
\includegraphics[clip, height=3cm]{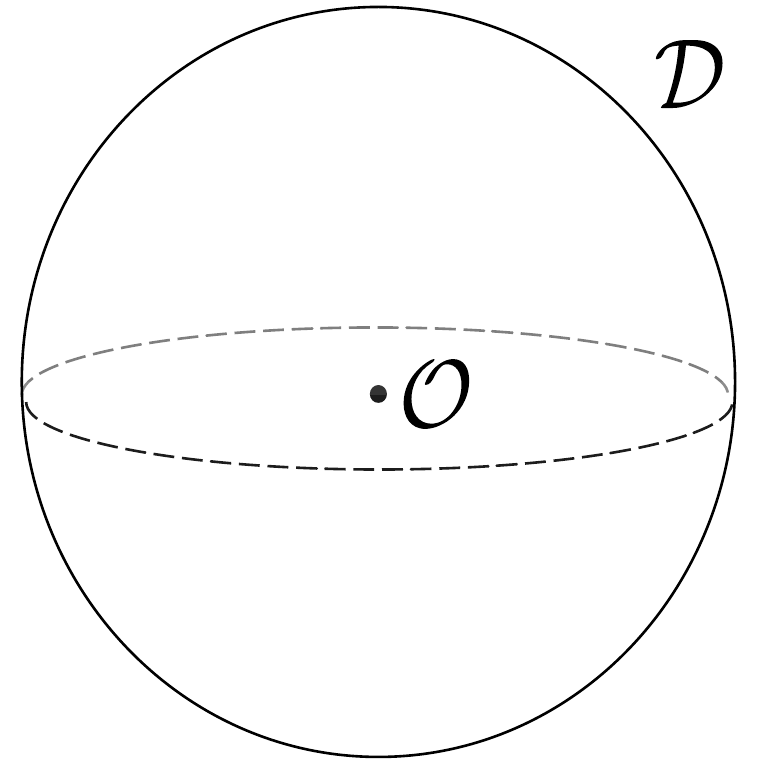}
\subcaption{Sphereical defect and local operator}
\end{minipage}
\begin{minipage}
{0.45\hsize}
\centering
\includegraphics[clip, height=3cm]{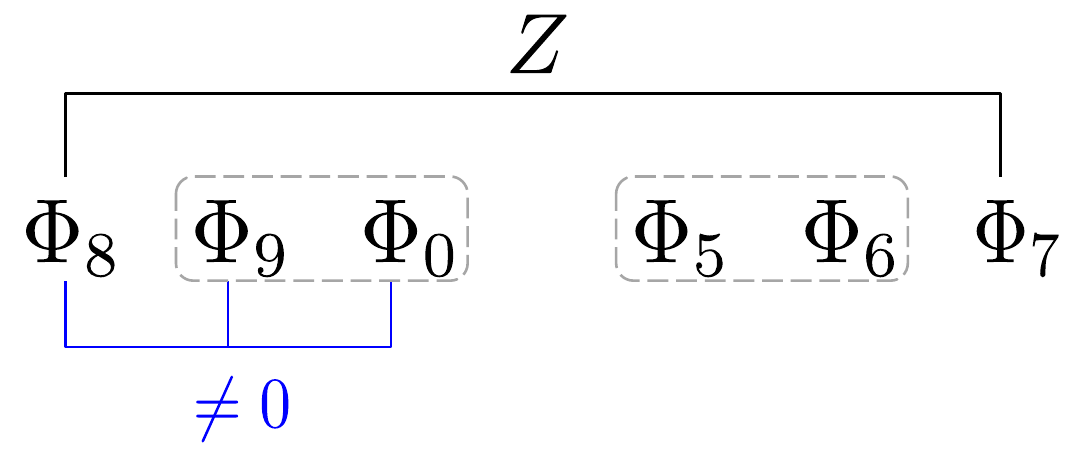}
\subcaption{Scalars and R-symmetry}
\end{minipage}
\caption{The half-BPS defect breaks $\mathfrak{psu}(2|2)^2$ down to $\mathfrak{su}(2|1)^2$. Their bosonic subgroups $\mathfrak{su}(2)^2$ and $\mathfrak{u}(1)^2$ are realized as follows: (a) A spherical defect and a local operator inserted at the origin preserves the $SO(4)$ Lorentz symmetry. (b) Among the six scalars in $\mathcal{N}=4$ SYM,  two of them $\Phi_{7,8}$ are used to define the vacuum of the dynamic spin chain, and three of them $\Phi_{8,9,0}$ acquire nontrivial vacuum expectation values in the presence of the defect. Therefore the residual symmetry group is $U(1)^2$ which rotate $\Phi_{9,0}$ and $\Phi_{5,6}$ (encircled by the dashed curves).}\label{fig:symmetrygroup}
\end{figure}

\paragraph{$\mathfrak{su}(2|1)^2$ symmetry}
In the presence of the half-BPS defect, the $\mathfrak{psu}(2|2)^2$ symmetry is broken down to a subgroup which is an intersection of $\mathfrak{psu}(2|2)^2$ and the defect superconformal symmetry $\mathfrak{osp}(4|4, \mathbb{R})$. To understand the structure of this subalgebra, it is useful to place the single-trace operator at the origin and consider a spherical defect of radius $r$ around it. As explained in Figure \ref{fig:symmetrygroup}, this configuration manifestly preserves the $SO(4) (\simeq SU(2)^2)$ rotation symmetry, and the $U(1)^2$ $R$-symmetry which rotates $\Phi_{9,0}$ and $\Phi_{5,6}$ in \eqref{eq:XiYidef}. Once the fermionic charges are included, these bosonic symmetries get completed into the $\mathfrak{su}(2|1)^2$ subalgebra whose generators are given by
\beq
\begin{aligned}
&\mathcal{R}\equiv R^{1}{}_{1}-R^{2}{}_{2}\comma\qquad &&L^{\alpha}{}_{\beta}\comma\qquad &&
\mathcal{S}^{a}{}_{\alpha} \equiv S^{a}{}_{\alpha}+ ir\epsilon_{\alpha\beta}\sigma^{ab}Q^{\beta}{}_{b}\comma\\
&\dot{\mathcal{R}}\equiv \dot{R}^{1}{}_{1}-\dot{R}^{2}{}_{2}\comma&&\dot{L}^{\dot{\alpha}}{}_{\dot{\beta}}\comma&&
\dot{\mathcal{S}}^{\dot{a}}{}_{\dot{\alpha}} \equiv  \dot{S}^{\dot{a}}{}_{\dot{\alpha}}+ ir\epsilon_{\dot{\alpha}\dot{\beta}}\sigma^{\dot{a}\dot{b}}\dot{Q}^{\dot{\beta}}{}_{\dot{b}}\comma
\comma
\end{aligned}
\eeq
where $r$ is the radius of the spherical defect and $\sigma^{ab}$ is a symmetric tensor defined by
\beq
\sigma^{11}=\sigma^{22}=0\comma \quad \sigma^{12}=\sigma^{21}=1\period
\eeq

These generators satisfy the following algebras:\fn{Since the left and the right $\mathfrak{su}(2|1)$ have the identical structure, here we only write down the commutation relations for the left part.} 
\beq
\begin{aligned}
&[L^{\alpha}{}_{\beta},\mathcal{R}]=0\comma\quad [L^{\alpha}{}_{\beta},\mathcal{S}^{a}_{\gamma}]=\delta^{\alpha}_{\gamma}\mathcal{S}^{a}_{\beta}-\frac{1}{2}\delta^{\alpha}_{\beta}\mathcal{S}^{a}_{\gamma}\comma\\
& [\mathcal{R},\mathcal{S}^{1}{}_{\alpha}]=+\mathcal{S}^{1}{}_{\alpha}\comma\quad [\mathcal{R},\mathcal{S}^{2}{}_{\alpha}]=-\mathcal{S}^{2}{}_{\alpha}\\
&\{\mathcal{S}^{a}{}_{\alpha},\mathcal{S}^{b}{}_{\beta}\} =\epsilon_{\alpha\beta}\epsilon^{ab}\left(K+r^2 P +ir \mathcal{R}\right)+i\sigma^{ab}\left( \epsilon_{\alpha\gamma}L^{\gamma}{}_{\beta}+\epsilon_{\beta\gamma}L^{\gamma}{}_{\alpha}\right)
\end{aligned}
\eeq
Note that the central charges $P$ and $K$ always appear in the combination $K+r^2 P +ir \mathcal{R}$, and the algebra is isomorphic to $\mathfrak{su}(2|1)$ without central extension.

We now discuss the implication of $\mathfrak{su}(2|1)^2$ symmetry on the two-particle overlap. Explicitly, we impose\footnote{The integrable boundary states with $\mathfrak{su}(2|1)^2$ were analyzed also in \cite{Hofman:2007xp}. Their boundary state is in the mirror channel while our boundary state is in the physical channel. Since the non-relativistic worldsheet theories in the mirror and the physical channels are inequivalent, there seems to be no simple relation between the two boundary states.}
\beq
\langle \mathcal{D}|j|\mathcal{X}^{A\dot{A}}(u)\mathcal{X}^{B\dot{B}}(\bar{u})\rangle=0\comma\qquad j\in \mathfrak{su}(2|1)_{L}\oplus \mathfrak{su}(2|1)_{R}\period
\eeq
Here $A,B$ and $\dot A,\dot B$ label abstractly the indices carried by the left and right excitations separately.

Owing to the structure of the symmetry algebra $\mathfrak{su}(2|1)^2$, it is useful to factorize the overlap into the left and the right parts as
\beq\label{eq:structuretwoparticle}
\langle \mathcal{D}|\mathcal{X}^{A\dot{A}}(u)\mathcal{X}^{B\dot{B}}(\bar{u})\rangle =F_0(u)\times \langle \mathfrak{d}|\chi^{A}(u)\chi^{B}(\bar{u})\rangle \times \langle \mathfrak{d}|\chi^{\dot{A}}(u)\chi^{\dot{B}}(\bar{u})\rangle 
\eeq
and discuss constraints from the left and the right $\mathfrak{su}(2|1)$'s separately. Here $F_0$ is an overall scalar factor which will be determined later.
\paragraph{Constraints from bosonic symmetry} Let us impose the invariance under the bosonic symmetry. First, from the left $SU(2)$ rotation symmetry, we can constrain the overlaps involving fermions as
\beq
\begin{aligned}
&\bra{\mathfrak{d}}\psi^{\alpha}(u)\psi^{\beta}(\bar{u})\rangle = \epsilon^{\alpha\beta}\comma\\
&\bra{\mathfrak{d}}\phi^{a}(u)\psi^{\beta}(\bar{u})\rangle =\bra{\mathfrak{d}}\psi^{\alpha}(u)\phi^{b}(\bar{u})\rangle =0\period
\end{aligned}
\eeq
To constrain the overlap of two bosons, we impose the invariance under the $U(1)$ generator $K+r^2 P +ir \mathcal{R}$. This is equivalent to imposing the invariance under $\mathcal{R}$ since the action of $K+r^{2}P$ annihilates a parity-invariant pair of excitations $|\mathcal{X}(u)\mathcal{X}(\bar{u})\rangle$ as can be verified from \eqref{eq:centralchargeaction}. We then get
\beq
\begin{aligned}
0&=\bra{\mathfrak{d}}\mathcal{R}\ket{\phi^{1}(u)\phi^{1}(\bar{u})}=2\langle\mathfrak{d}\ket{\phi^{1}(u)\phi^{1}(\bar{u})}\period\\
0&=\bra{\mathfrak{d}}\mathcal{R}\ket{\phi^{2}(u)\phi^{2}(\bar{u})}=-2\langle\mathfrak{d}\ket{\phi^{2}(u)\phi^{2}(\bar{u})}\period
\end{aligned}
\eeq
Therefore, only nonzero components are the following ones:
\beq\label{eq:defkpkm}
\begin{aligned}
&\bra{\mathfrak{d}}\phi^{1}(u)\phi^{2}(\bar{u})\rangle =: k_+ (u)\comma \quad \bra{\mathfrak{d}}\phi^{2}(u)\phi^{1}(\bar{u})\rangle =: k_- (u)\period
\end{aligned}
\eeq
\paragraph{Rule of pulling out $Z$ marker} In order to analyze constraints from fermionic generators, we need to impose {\it by hand} a rule of pulling out $Z$ markers from the closed-string state. Roughly speaking this determines how an insertion and a removal of $Z$ field change the one-point function, and it is therefore related to the expectation value of the scalar field. For now, we simply assume the following rule
\begin{shaded}
\beq\label{eq:pullingout}
\langle \mathcal{D}|Z\mathcal{X}\rangle =\frac{x_s}{r} \langle \mathcal{D}|\mathcal{X}\rangle\comma
\eeq
\end{shaded}
\noindent without specifying the value of $x_s(=x(is))$. Note that we multiplied a factor $1/r$ to account for the mass dimension of $Z$. 

The value of $x_s$ has a clear physical meaning both in the gauge theory and string theory. On the gauge-theory side, it can be interpreted as an expectation value of the $Z$ field in the classical background sourced by the interface defect. On the string-theory side, it is a parameter which distinguishes different boundary states and is related to a momentum carried by the boundary state as we discuss in more detail in section \ref{subsec:excitedB}. There we also show that one can change the value of $x_s$ by considering a bound state of the boundary state and a bulk particle. This feature turns out to be essential in order to reproduce the results obtained at weak coupling \cite{Buhl-Mortensen:2017ind}.

\paragraph{Constraints from fermionic symmetry}
Next we study constraints from the fermionic symmetry. The fermionic symmetry exchanges bosons and fermions in the dynamic spin chain and thereby allows us to determine $k_{+}$ and $k_{-}$ in \eqref{eq:defkpkm}.

Let us first consider the state $|\phi^{a}(u)\psi^{\alpha}(\bar{u})\rangle$. The action of $\mathcal{S}$ reads
\begin{align}
\mathcal{S}^{b}_{\beta}|\phi^{a}(u)\psi^{\alpha}(\bar{u})\rangle=&{\sf c}\epsilon^{ba}\epsilon_{\beta\gamma}|Z^{-}\psi^{\gamma}(u)\psi^{\alpha}(\bar{u})\rangle+\bar{{\sf d}}\delta_{\beta}^{\alpha}|\phi^{a}(u)\phi^{b}(\bar{u})\rangle\nonumber\\
&+i{\sf a} r \epsilon_{\beta\gamma}\sigma^{bc}\delta^{a}_{c}|\psi^{\gamma}(u)\psi^{\alpha}(\bar{u})\rangle+i \bar{{\sf b}}r \epsilon_{\beta\gamma}\sigma^{bc}\epsilon^{\gamma\alpha}\epsilon_{cd}|\phi^{a}(u)Z\phi^{d}(\bar{u})\rangle\,,\\
=&r\epsilon^{ba}\epsilon_{\beta\gamma}\left(\frac{{\sf c}}{x_s}+i{\sf a} s^{b}\right)|\psi^{\gamma}(u)\psi^{\alpha} (\bar{u})\rangle+\delta^{\alpha}_{\beta}\left(\bar{{\sf d}}+ix_s \bar{{\sf b}} s^{b}\frac{x^{+}}{x^{-}}\right)|\phi^{a}(u)\phi^{b}(\bar{u})\rangle\comma\nonumber
\end{align}
where $\bar{\sf b}$ and $\bar{\sf d}$ are given by ${\sf b}(\bar{u})$ and ${\sf d}(\bar{u})$, and $s^{b}$ is defined by $s^{1}=-s^{2}=1$.
Note that here we have already used the rule \eqref{eq:pullingout} to pull out $Z$ markers.

By contracting the state against the boundary state $\langle \mathfrak{d}|$, we obtain
\beq
\begin{aligned}
&0=\langle \mathfrak{d}|\mathcal{S}^{b}_{\beta}|\phi^{a}(u)\psi^{\alpha}(\bar{u})\rangle\\
&\iff \begin{cases}0=r \delta_{\beta}^{\alpha}\left(\frac{{\sf c}}{x_s}+i{\sf a} \right)+\delta^{\alpha}_{\beta}\left(\bar{{\sf d}}+ix_s \bar{{\sf b}} \frac{x^{+}}{x^{-}}\right)k_{-}(u)&(a=2, b=1)\,,\\
0=-r \delta_{\beta}^{\alpha}\left(\frac{{\sf c}}{x_s}-i{\sf a} \right)+\delta^{\alpha}_{\beta}\left(\bar{{\sf d}}-ix_s \bar{{\sf b}} \frac{x^{+}}{x^{-}}\right)k_{+}(u)&(a=1, b=2)\,.\end{cases}
\end{aligned}
\eeq
Solving these equations, we get
\beq\label{eq:kpkmsolution}
\begin{aligned}
k_{+}(u)&=-ir \frac{1-\frac{1}{x^{+}x_s}}{1+\frac{x_s}{x^{-}}}\comma\qquad
&&k_{-}(u)=-ir \frac{1+\frac{1}{x^{+}x_s}}{1-\frac{x_s}{x^{-}}}\period
\end{aligned}
\eeq
This determines the two-particle overlaps \eqref{eq:structuretwoparticle} up to an overall factor $F_0(u)$.

\subsection{Constraining the scalar factor}
Having determined the matrix structure of the two-particle overlap, we next determine $F_0(u)$  by imposing a set of consistency conditions: Watson's equation and the crossing symmetry. We also check that our solution satisfies the boundary Yang-Baxter equations.
\paragraph{String frame and redefinition of the overall factor} To impose these constraints, it is more convenient to change the definition of the bosonic excitation in the dynamic spin chain as
\beq
\phi_{\rm new}=Z^{1/4} \phi_{\rm old}Z^{1/4}\period
\eeq 
The excitations defined in this way are called {\it string-frame} excitations \cite{Arutyunov:2006yd} while the original ones are called {\it spin-chain frame} excitations. The spin-chain frame is more convenient for the comparison with the gauge theory while the string-frame is more natural for the analysis of the dual string worldsheet. After this change, the two-particle overlap of two scalar gets modified as follows:
\beq
\begin{aligned}
\left.\langle \mathfrak{d} | \phi^{a}(u)\phi^{b}(\bar{u})\rangle \right|_{\rm string}&= \left.\langle \mathfrak{d} |Z^{1/4}\phi^{a}(u)Z^{1/2}\phi^{b}(\bar{u})Z^{1/4}\rangle\right|_{\rm spin}\,,\\
&=\frac{x_s }{r}\sqrt{\frac{x^{+}}{x^{-}}}\left.\langle \mathfrak{d} |\phi^{a}(u)\phi^{b}(\bar{u})\rangle\right|_{\rm spin}\,.
\end{aligned}
\eeq

\begin{figure}[t]
 \centering
 \begin{minipage}{0.45\hsize}
 \includegraphics[clip, height=2cm]{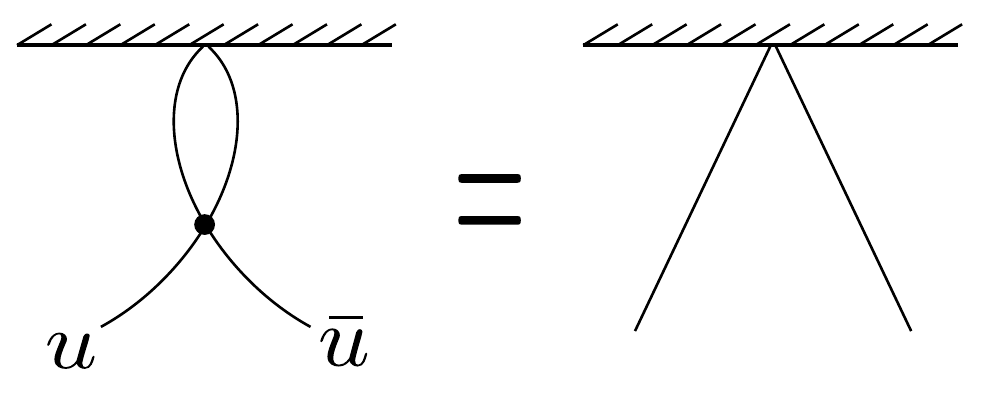}
 \subcaption{Watson's equation}\label{subfig:watson}
 \end{minipage}
 \begin{minipage}{0.45\hsize}
 \includegraphics[clip,width=7cm, height=2cm]{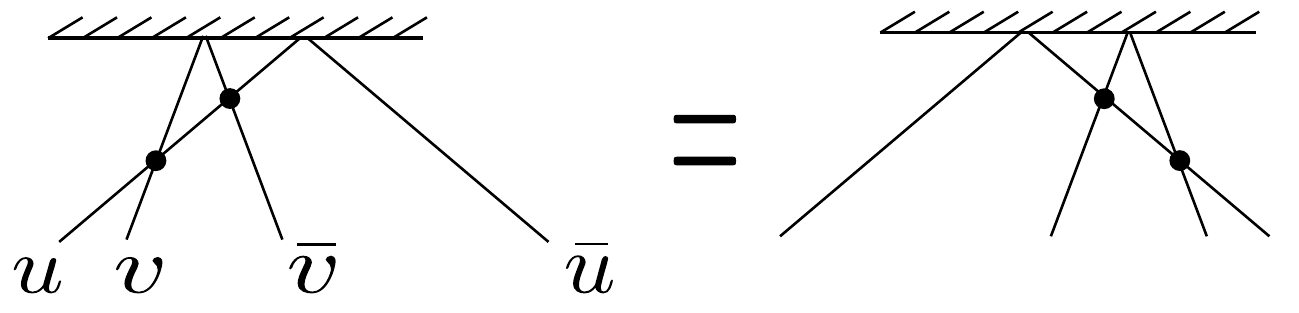}
 \subcaption{Boundary Yang-Baxter equation}\label{subfig:bYB}
 \end{minipage}
 \caption{The Watson's equation and boundary Yang-Baxter equations. Here $u$, $v$, $\bar{u}$ and $\bar{v}$ are rapidities of excitations and their parity conjugates. The black dots in the figures denote the S-matrix of the excitations.}
 \end{figure}

For later purposes, we rescale the overall factor (from $F_0$ to $F$) and write the two-particle overlaps in the string frame as follows\footnote{After this paper came out on arXiv, the paper \cite{Gombor:2020kgu} appeared in which they studied the matrix structures of two-particle overlaps corresponding to integrable boundary states in the dynamic $\mathfrak{psu}(2|2)$ spin chain, and obtained a result which seems consistent with \eqref{eq:matrixstructurestringframe}.}:
\begin{shaded}
\begin{align}\label{eq:matrixstructurestringframe}
\begin{aligned}
\langle \mathcal{D}|\mathcal{X}^{A\dot{A}}(u)\mathcal{X}^{B\dot{B}}(\bar{u})\rangle &=F(u)\times \langle \mathfrak{d}|\chi^{A}(u)\chi^{B}(\bar{u})\rangle \times \langle \mathfrak{d}|\chi^{\dot{A}}(u)\chi^{\dot{B}}(\bar{u})\rangle \comma\\
\bra{\mathfrak{d}}\psi^{\alpha}(u)\psi^{\beta}(\bar{u})\rangle &=\frac{x_s^2-(x^{-})^2}{x^{-}(x^{+}+x^{-})} \epsilon^{\alpha\beta}\comma\\
\bra{\mathfrak{d}}\phi^{1}(u)\phi^{2}(\bar{u})\rangle &= -i x_s\frac{x_s^2-(x^{-})^2}{x^{-}(x^{+}+x^{-})}\sqrt{\frac{x^{+}}{x^{-}}}\frac{1-\frac{1}{x^{+}x_s}}{1+\frac{x_s}{x^{-}}}\comma\\
\bra{\mathfrak{d}}\phi^{2}(u)\phi^{1}(\bar{u})\rangle &= -i x_s\frac{x_s^2-(x^{-})^2}{x^{-}(x^{+}+x^{-})}\sqrt{\frac{x^{+}}{x^{-}}}\frac{1+\frac{1}{x^{+}x_s}}{1-\frac{x_s}{x^{-}}}\period
\end{aligned}
\end{align}
\end{shaded}
\paragraph{Watson's equation}
The first consistency condition is Watson's equation, which states that one needs to multiply the S-matrix in order to reorder excitations, see Figure \ref{subfig:watson}. Written explicitly, it says
\beq
\langle\mathcal{D}|\mathbb{S}|\mathcal{X}_1\mathcal{X}_2\rangle=\langle \mathcal{D}|\mathcal{X}_1\mathcal{X}_2\rangle\period
\eeq 
Note that this is a matrix-valued equation and therefore is an over-constrained system for a single scalar factor $F(u)$. 
Remarkably we found that it reduces to a single constraint on $F$,
\beq
\frac{F (u)}{F (\bar{u})}=S_0 (u,\bar{u})\comma
\eeq
where $S_0$ is the overall scalar factor for the bulk S-matrix, given by 
\begin{align}
S_0 (u_1,u_2)&=\frac{x_1^{+}-x_2^{-}}{x_1^{-}-x_{2}^{+}}\frac{1-1/x_1^{-}x_2^{+}}{1-1/x^{+}_1x_2^{-}}\frac{1}{\sigma^2 (u_1,u_2)}\comma
\end{align}
where $x_{1,2}\equiv x(u_{1,2})$ and $\sigma (u_1,u_2)$ is the dressing phase determined in \cite{Beisert:2006ez}.
 The equation can be solved by the following ansatz,
 \begin{shaded}
 \beq\label{eq:factorizinggu}
 F(u)=\frac{(x^{+}+x^{-})^3}{2(x_s-x^{+})(x_s+x^{-})(1-1/(x_sx^{+}))(1+1/(x_sx^{-}))}\frac{1+1/(x^{-})^2}{x^{-}+1/x^{+}}\frac{\sigma_B (u)}{\sigma (u,\bar{u})}\comma
 \eeq
 \end{shaded}
 \noindent
 where $\sigma_B (u)$ is an undetermined prefactor which we call the {\it boundary dressing phase}. At this point, $\sigma_B$ can be arbitrary as long as it satisfies  
 \beq
 \sigma_B (u)=\sigma_B (\bar{u})\period
 \eeq
 
 \paragraph{Boundary Yang-Baxter equation} We next consider the boundary Yang-Baxter equations as depicted in Figure \ref{subfig:bYB}:
\beq
\langle \mathcal{D}|\mathbb{S}_{24}\mathbb{S}_{34}|\mathcal{X}_1 (u)\mathcal{X}_2 (v)\mathcal{X}_{3}(\bar{v})\mathcal{X}_4(\bar{u})\rangle=\langle \mathcal{D}|\mathbb{S}_{13}\mathbb{S}_{24}|\mathcal{X}_1 (u)\mathcal{X}_2 (v)\mathcal{X}_{3}(\bar{v})\mathcal{X}_4(\bar{u})\rangle\period
\eeq
Here $\mathbb{S}_{ij}$ is the S-matrix between $\mathcal{X}_i$ and $\mathcal{X}_j$. Checking these equations is a straightforward yet tedious task. We verified that these equations are satisfied by the solution \eqref{eq:kpkmsolution}, regardless of the values of $F(u)$ and $x_s$. This provides further support for our assumption that the boundary state $|\mathcal{D}\rangle$ is an integrable boundary state. 

\begin{figure}[t]
 \centering
 \includegraphics[clip, height=2cm]{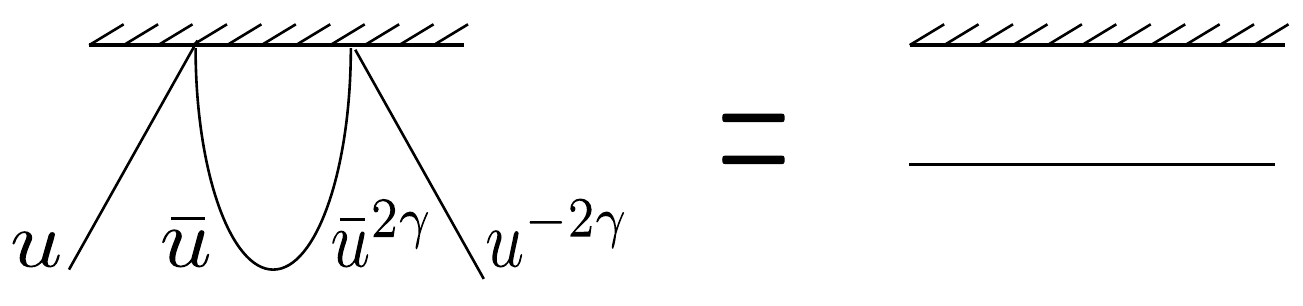}
 \caption{The crossing equation for the two-particle overlap. It requires that the overlap for the singlet states (shown in the figure) is trivial. In general, to write down the crossing equation, one needs to sum over all possible states that form a single pair $\mathcal{X}(\bar{u})$ and $\mathcal{X}(\bar{u}^{2\gamma})$. (This is why the middle two particles in the figure are connected by the curve.) In the case at hand, by judiciously choosing the external states, it reduces to a simple constraint given in \eqref{eq:crossing}.}\label{fig:crossing}
 \end{figure}

\paragraph{Crossing equation}
The third constraint is the crossing equation, which requires the overlap to be trivial when excitations form a singlet state of the centrally-extended $\mathfrak{psu}(2|2)^2$ symmetry. The singlet state has the same quantum numbers as the vacuum state of the dynamic spin chain and can be pair-created from the vacuum. Therefore imposing the crossing equation is physically equivalent to requiring the overlap to be invariant under the vacuum fluctuations. For more detailed discussions, see sections 6 and 7 of \cite{Jiang:2019xdz} (and also \cite{Ghoshal:1993tm}). In our case, it boils down to the following relation (see also Figure \ref{fig:crossing}):
\beq\label{eq:crossing}
\langle \mathcal{D}|D^{1\dot{1}} (u)D^{2\dot{2}}(\bar{u})D^{1\dot{1}}(\bar{u}^{2\gamma})D^{2\dot{2}}(u^{-2\gamma})\rangle=1\period
\eeq

To compute the left hand side, one has to analyze the four-particle form overlap. Here we use the assumption that the system is described by an integrable boundary state and factorize the four-particle overlap into a product of two-particle overlaps.  Under this assumption, \eqref{eq:crossing} becomes
\beq
\underbrace{\left[F(u)\left(\frac{x_s^2-(x^{-})^2}{x^{-}(x^{+}+x^{-})}\right)^2\right]}_{D^{1\dot{1}}(u)D^{2\dot{2}}(\bar{u})}\,\,\underbrace{\left[F(\bar{u}^{2\gamma})\left(\frac{x^{-}(1-x_s^2(x^{+})^2)}{(x^{+}+x^{-})}\right)^2\right]}_{D^{1\dot{1}}(\bar{u}^{2\gamma})D^{2\dot{2}}(u^{-2\gamma})}=1\period
\eeq
Substituting the expression \eqref{eq:factorizinggu} and using the identity\fn{This relation can be derived from the crossing equation for the dressing phase $\sigma (u,v)$:
\beq
\sigma (u_1,u_2)\sigma (u_1^{2\gamma},u_2)=\frac{(1-1/x_1^{+}x_2^{+})(1-x_1^{-}/x_2^{+})}{(1-1/x_1^{+}x_2^{-})(1-x_1^{-}/x_2^{-})}\period
\eeq
For details of the derivation, see Appendix G of \cite{Jiang:2019xdz}.
}
\beq
\frac{1}{\sigma (u,\bar{u})\sigma (\bar{u}^{2\gamma},u^{-2\gamma})}=\frac{4 (x^{-})^4}{(x^{+}+1/x^{+})(x^{-}+1/x^{-})}\left(\frac{1+\frac{1}{x^{+}x^{-}}}{x^{+}+x^{-}}\right)^{2}\comma
\eeq
we can rewrite the crossing equation as follows:
\beq
\sigma_B (u)\sigma_B(\bar{u}^{2\gamma})=\frac{1}{x_s^4}\frac{(x^{+}-x_s)^2(x^{-}+1/x_s)^2}{(x^{+}+1/x_s)^2(x^{-}-x_s)^2}\period
\eeq
This can be further rewritten using $\sigma_B(\bar{u})=\sigma_B (u)$ as
\beq\label{eq:boundarycross}
\sigma_B (u)\sigma_B (u^{2\gamma})=\frac{1}{x_s^{4}}\frac{(x^{+}-1/x_s)^2(x^{-}+x_s)^2}{(x^{+}+x_s)^2(x^{-}-1/x_s)^2}\period
\eeq

\subsection{Solving the crossing equation}
We now solve the functional equation \eqref{eq:boundarycross} by applying standard procedures explained in the review \cite{Vieira:2010kb}.  

\paragraph{Minimal solution} We first perform the mirror transformation and rewrite \eqref{eq:boundarycross} as
\beq\label{eq:aftermirror}
\sigma_B (u^{-\gamma})\sigma_B (u^{\gamma})=\frac{(x^{+}-1/x_s)^2(x^{-}+1/x_s)^2}{(x^{+}+x_s)^2(x^{-}-x_s)^2}\period
\eeq
We then consider the following ansatz (to be called the minimal solution)
\beq
\sigma^{\rm min}_B (u)=\frac{G(x^{-},x_s)G(x^{+},-1/x_s)}{G(x^{+},x_s)G(x^{-},-1/x_s)}\comma
\eeq
and write \eqref{eq:aftermirror} as
\beq
\frac{\left(G(x,-1/x_s)G(1/x,-1/x_s)\right)^{{\rm D}-{\rm D}^{-1}}}{\left(G(x,x_s)G(1/x,x_s)\right)^{{\rm D}-{\rm D}^{-1}}}=\frac{(x-1/x_s)^{2{\rm D}}(x+1/x_s)^{2{\rm D}^{-1}}}{(x+x_s)^{2{\rm D}}(x-x_s)^{2{\rm D}^{-1}}}\comma
\eeq
where ${\rm D}$ is the shift operator ${\rm D}\equiv\frac{i}{2}\del_u$. Solutions to this equation can be obtained by dealing with a simpler equation,
\beq\label{eq:morebasic}
G(x,y)G(1/x,y)=\left(\frac{x-\frac{1}{y}}{\sqrt{x}}\right)^{\frac{-2{\rm D}}{{\rm D}-{\rm D}^{-1}}}\left(\frac{x+\frac{1}{y}}{\sqrt{x}}\right)^{\frac{-2{\rm D}^{-1}}{{\rm D}-{\rm D}^{-1}}}\comma
\eeq
which implies
\beq\label{eq:crossingintermediate}
\begin{aligned}
G(x,y)G(1/x,y)G(x,1/y)G(1/x,1/y)&=(u-v)^{\frac{-2{\rm D}}{{\rm D}-{\rm D}^{-1}}}(u+v)^{\frac{-2{\rm D}^{-1}}{{\rm D}-{\rm D}^{-1}}}\\
&=\left(\frac{\Gamma[1+i(u+v)]}{\Gamma[1-i (u-v)]}\right)^2\comma
\end{aligned}
\eeq
with $v\equiv g (y+1/y)$ (or equivalently $y=x(v)$). In the second equality of \eqref{eq:crossingintermediate}, we  have used a series-expansions of the exponents
\beq
\begin{aligned}
&\frac{-2{\rm D}}{{\rm D}-{\rm D}^{-1}}=\frac{2{\rm D}^2}{1-{\rm D}^2}=2\sum_{n=1}^{\infty}{\rm D}^{2n}\comma\\
&\frac{-2{\rm D}^{-1}}{{\rm D}-{\rm D}^{-1}}=-\frac{2{\rm D}^{-2}}{1-{\rm D}^{-2}}=-2\sum_{n=1}^{\infty}{\rm D}^{-2n}\period
\end{aligned}
\eeq

The functional equation \eqref{eq:crossingintermediate} can be solved by applying standard techniques of the Riemann-Hilbert problem (see e.g.~\cite{Vieira:2010kb}). The result reads
\begin{shaded}
\beq\label{eq:solutionGxy}
\frac{1}{i}\log G(x,y)=\frac{2}{i}\oint_{|z|=1}\frac{dz}{2\pi i}\oint_{|w|=1}\frac{dw}{2\pi i}\frac{1}{x-z}\frac{1}{y-w}\log \mathfrak{G}(z,w)\comma
\eeq 
with
\beq
\mathfrak{G}(z,w)=\frac{\Gamma[1+ig(z+\tfrac{1}{z}+w+\tfrac{1}{w})]}{\Gamma[1-i g(z+\tfrac{1}{z}-w-\tfrac{1}{w})]}\period
\eeq
\end{shaded}
\noindent
The solution \eqref{eq:solutionGxy} is valid for $|x|>1$ and $|y|>1$, and the result for other parameter regions can be obtained by   analytic continuation.
\paragraph{CDD ambiguity} Using the minimal solution $\sigma^{\rm min}(u)$, we can construct infinitely many solutions to the crossing equation by multiplying a factor $\sigma_{\rm CDD}(u)$ which satisfies
\beq\label{eq:CDDrel}
\sigma_{\rm CDD}(u)=\sigma_{\rm CDD}(\bar{u})\comma\qquad \sigma_{\rm CDD}(u)\sigma_{\rm CDD}(u^{2\gamma})=1\period
\eeq
This is an analogue of the Castillejo-Dalitz-Dyson (CDD) ambiguity \cite{Castillejo:1955ed} for the bulk S-matrix. In particular, for any odd function of the magnon energy $f_{\rm odd}(E)$, one can verify that $\sigma_{\rm CDD}(u)=e^{f_{\rm odd}(E)}$ satisfies the relations \eqref{eq:CDDrel}.

As we see later, the choice that reproduces the weak-coupling results in the literature turns out to be $\sigma_{\rm CDD}(u)=2^{-4E(u)}$. 
Thus our proposal for the solution to the crossing equation relevant for the D5-brane interface is
\begin{shaded}
\beq\label{eq:crossingsolfinal}
\sigma_B (u)=2^{-4E(u)}\frac{G(x^{-},x_s)G(x^{+},-1/x_s)}{G(x^{+},x_s)G(x^{-},-1/x_s)}\period
\eeq  
\end{shaded}
\noindent 

 Let us now make one remark: The same CDD factor appeared in the analysis of structure constants of determinant operators in \cite{Jiang:2019xdz,Jiang:2019zig}, and it was interpreted\footnote{See section 7.5 of \cite{Jiang:2019xdz}.} as an extra spacetime dependence associated to a conformal transformation which maps the symmetric configuration (in which operators are at $-1$, $0$ and $1$ along a line) to the canonical configuration (in which operators are at $0$, $1$ and $\infty$).  A similar argument holds also for the defect one-point function. Normally we consider a planar defect and define the structure constant $C_{\mathcal{O}}$ by
\beq
\langle \mathcal{O}\rangle_{\mathcal{D}}=\frac{C_{\mathcal{O}}}{x_{\perp}^{\Delta}}\comma
\eeq
where $x_{\perp}$ is the distance between the defect and the operator. However, to apply integrability, it is more convenient to use a spherical defect with a unit radius and define the structure constant (to be denoted by $C_{\mathcal{O}}^{\rm sphere}$) as the one-point function in that configuration. Analyzing the conformal (and the $R$-symmetry) transformations which map the two configurations, we find that the relation between the two structure constants is given by\footnote{Here we have been suppressing the dependence on the $R$-symmetry polarizations. To reproduce the factor $2^{-(\Delta-J)}$ in \eqref{eq:relationtwostructure}, we also need to keep track of the $R$-symmetry polarizations. See \cite{Jiang:2019xdz} for details.}
\beq\label{eq:relationtwostructure}
C_{\mathcal{O}}=2^{-(\Delta-J)}C^{\rm sphere}_{\mathcal{O}}\period
\eeq
The factor $2^{-(\Delta-J)}$ is precisely the origin of the CDD factor in \eqref{eq:crossingsolfinal}. In other words, if we define the structure constant using the spherical defect, we would not need the CDD factor.
\paragraph{Weak coupling expansions} We now expand the boundary dressing phase $\sigma_B$ at weak coupling in order to perform a comparison with the literature. To do so we assume that the absolute value of $x_s$ is larger than $1$: $|x_s|> 1$. The assumption is justified eventually by the match with the weak-coupling results in the literature. 

Since $\sigma_B$ is given by \eqref{eq:crossingsolfinal}, we need expansions of both $G(x,y)$ and $G(x,1/y)$ with $|x|>1$ and $|y|>1$. Let us first discuss the expansion of $G(x,y)$. This can be done by expanding \eqref{eq:solutionGxy} in a power series of $1/x$ and $1/y$ as
\beq\label{eq:Gxyexpanded}
\frac{\log G(x,y)}{i}=\sum_{r,s=1}^{\infty}\frac{c_{r,s}}{x^{r}y^{s}}\comma
\eeq
with
\beq\label{eq:crsdef}
c_{r,s}\equiv\frac{2}{i}\oint_{|z|=1}\frac{dz z^{r-1}}{2\pi i}\oint_{|w|=1}\frac{dw w^{s-1}}{2\pi i}\log \mathfrak{G}(z,w)\period
\eeq
We then use the integral representation of $\log \Gamma$,
\beq\label{eq:loggammaint}
\log \Gamma (z)=\int_{0}^{\infty} \frac{dt}{t}e^{-t}\left(z-1-\frac{1-e^{-t (z-1)}}{1-e^{-t}}\right)\comma
\eeq
and rewrite \eqref{eq:crsdef} as
\beq
c_{r,s}=\frac{2}{i}\int_{0}^{\infty}\frac{dt}{t}\oint_{|z|=1}\frac{dz z^{r-1}}{2\pi i}\oint_{|w|=1}\frac{dw w^{s-1}}{2\pi i}\frac{e^{-i g t(z+\frac{1}{z}+w+\frac{1}{w})}-e^{i g t (z+\frac{1}{z}-w-\frac{1}{w})}}{e^{t}-1}\period
\eeq
To proceed we expand the integrand using
\beq
e^{i a (b+\frac{1}{b})}=\sum_{k=-\infty}^{\infty}i^k b^k J_k (2a)\comma
\eeq
and perform the integrals of $z$ and $w$. As a result we obtain
\beq
c_{r,s}=\frac{2(1-(-1)^r)}{i^{r+s+1}}\int_{0}^{\infty}dt\frac{ J_r (2gt)J_s (2gt)}{t(e^{t}-1)}\comma
\eeq
where $J$'s are the Bessel functions. Expanding the integrand in powers of $g$ and performing the integral, we get
\beq\label{eq:seriescrs}
c_{r,s}=\sum_{n=0}^{\infty}g^{r+s+2n}c_{r,s}^{(n)}\comma
\eeq
where
\beq
c_{r,s}^{(n)}=2 (-1)^{n}\frac{(1-(-1)^{r})}{i^{r+s+1}}\frac{(2n+r+s-1)!(2n+r+s)!}{n!(n+r)!(n+s)!(n+r+s)!}\zeta_{2n+r+s}\comma
\eeq
with $\zeta_n$ being the zeta function. From \eqref{eq:Gxyexpanded} and \eqref{eq:seriescrs}, one can show $G(x,y)=1+\mathcal{O}(g^4)$ at weak coupling.

We next consider $G(x,1/y)$. Since the magnitude of the second argument is less than $1$ ($|1/y|<1$), one has to analytically continue the integral representation \eqref{eq:solutionGxy}. Upon doing so, the integral picks up a contribution from a pole at $w=1/y$. As a result we have
\beq
\frac{1}{i}\log G(x,1/y)=\chi_{\rm int}(x,y)+\chi_{\rm pole}(x,y)\comma
\eeq 
with
\beq
\begin{aligned}
\chi_{\rm int}(x,y)&\equiv\frac{2}{i}\oint_{|z|=1}\frac{dz}{2\pi i}\oint_{|w|=1}\frac{dw}{2\pi i}\frac{1}{x-z}\frac{1}{\frac{1}{y}-w}\log \mathfrak{G}(z,w)\comma\\
\chi_{\rm pole}(x,y)&\equiv\frac{2}{i}\oint_{|z|=1}\frac{dz}{2\pi i}\frac{1}{x-z}\log \mathfrak{G}(z,y)\period
\end{aligned}
\eeq
By rewriting the integrand, we can further decompose $\chi_{\rm int}$ into two terms
\beq
\chi_{\rm int}(x,y)=-\frac{1}{i}\log G(x,y)-\bar{\chi}_{\rm int}(x)\comma
\eeq
with
\beq
\bar{\chi}_{\rm int}(x)\equiv\frac{2}{i}\oint_{|z|=1}\frac{dz}{2\pi i}\oint_{|w|=1}\frac{dw}{2\pi i}\frac{1}{x-z}\frac{1}{w}\log \mathfrak{G}(z,w)\period
\eeq
As we already know the expansion of $G(x,y)$, the remaining tasks are to expand $\chi_{\rm pole}$ and $\bar{\chi}_{\rm int}$. Using the integral representation of $\log \Gamma$ \eqref{eq:loggammaint}, we get 
\beq
\begin{aligned}
\chi_{\rm pole}(x,y)&=\sum_{r=1}^{\infty}\frac{d_r}{x^{r}}\comma\qquad
\bar{\chi}_{\rm int}(x)=\sum_{r=1}^{\infty}\frac{e_r}{x^{r}}\comma
\end{aligned}
\eeq
where $d_k$ and $e_k$ are given by
\beq
\begin{aligned}
d_r&=4g \delta_{r,1}\int_0^{\infty} dt \frac{e^{-t}}{t}+\frac{2(1-(-1)^r)}{i^{r+1}}\int_{0}^{\infty}\frac{e^{-it v}J_r (2g t)}{t(e^t -1)}\comma\\
e_r&=4g \delta_{r,1}\int_0^{\infty} dt \frac{e^{-t}}{t}+\frac{2(1-(-1)^{r})}{i^{r+1}}\int_{0}^{\infty}dt \frac{J_r(2gt)J_0 (2gt)}{t(e^{t}-1)}\comma
\end{aligned}
\eeq
where $v$ is the rapidity for $y$, $v= g(y+1/y)$.
Expanding these integrals in powers of $g$, we get
\beq
d_r=\sum_{n=0}^{\infty} g^{r+2n}d_{r}^{(n)}\comma\qquad 
e_r=\sum_{n=0}^{\infty} g^{r+2n}e_{k}^{(n)}\comma
\eeq
with
\beq
\begin{aligned}
d_r^{(n)}&=\frac{2(1-(-1)^{r})(-1)^{n+1}}{n! (n+r)! i^{r+1}}\Psi^{(2n+r-1)}(1+i v)\comma\\
e_{r}^{(n)}&=\begin{cases}-4 \gamma_{E}\qquad & (r=1, n=0)\,,\\ 2 (-1)^{n}\frac{(1-(-1)^{r})}{i^{r+1}}\frac{(2n+r-1)!(2n+r)!}{\left(n!(n+r)!\right)^2}\zeta_{2n+r}\qquad &({\rm others})\,.\end{cases} 
\end{aligned}
\eeq
Here $\Psi^{(n)}$ is the $n$-th derivative of the Euler digamma function and $\gamma_{E}$ is the Euler-Mascheroni constant. Therefore $\log G(x,1/y)$ can be expanded up to one loop as
\beq
\frac{1}{i}\log G (x,1/y)=\frac{4g^2}{u}\left(\Psi (1+i v)+\gamma_{E}\right)+\mathcal{O}(g^4)\period
\eeq

Putting together everything, we get the following expansion of the boundary dressing phase:
\beq\label{eq:sigmaBexpanded}
\sigma_B (u)=\frac{1}{4}\left[1+\frac{4g^2}{u^{2}+\frac{1}{4}}\left(\Psi (1+s)+\gamma_E-\log 2\right)+\mathcal{O}(g^4)\right]\period
\eeq
Already at this stage, it is worth pointing out that the one-loop result in \eqref{eq:sigmaBexpanded} takes the same form as the ``flux factor'' in \cite{Buhl-Mortensen:2017ind} if we set $s=(k-1)/2$. In the following subsections, we make this heuristic observation into a more concrete statement and show that our results are indeed in agreement with \cite{Buhl-Mortensen:2017ind}. 
\subsection{Excited boundary states and comparison with perturbation theory\label{subsec:excitedB}}
Having solved the crossing equation, we now make a comparison with the results from perturbation theory. We focus on the so-called $SU(2)$ sector (see below), for which the one-loop results are available \cite{deLeeuw:2015hxa,Buhl-Mortensen:2015gfd,Buhl-Mortensen:2017ind}. We leave the comparison in other sectors to a future work.
\paragraph{Two-particle overlap in $SU(2)$ sector} In $\mathcal{N}=4$ SYM, we can define subsectors of operators in which the action of the dilatation operator is closed at all orders in perturbation theory. The simplest subsector is the so-called $SU(2)$ sector, which consists of operators made out of two complex scalars. In order to make contact with \cite{deLeeuw:2015hxa,Buhl-Mortensen:2015gfd,Buhl-Mortensen:2017ind}, we choose the two scalars to be $Z$ and $\tilde{\Phi}$ with\footnote{Normally we simply choose $Z$ and $\Phi^{1\dot{1}}$ to define the $SU(2)$ sector. However, the one-point functions in that sector vanish owing to the matrix structure of the overlap \eqref{eq:matrixstructurestringframe}, and they do not correspond to the setup discussed in \cite{deLeeuw:2015hxa,Buhl-Mortensen:2015gfd,Buhl-Mortensen:2017ind}. This is the reason  why we have chosen a particular linear combination given in \eqref{eq:defphitil}.}
\beq\label{eq:defphitil}
\tilde{\Phi}\equiv\frac{\Phi^{1\dot{1}}+\Phi^{2\dot{2}}+\Phi^{1\dot{2}}+\Phi^{2\dot{1}}}{2}\period
\eeq
Then the two-particle overlap in this sector, $f_{SU(2)}(u)\equiv\langle \mathcal{D}|\tilde{\Phi}(u)\tilde{\Phi}(\bar{u})\rangle$, is given by
 \beq\label{eq:fsu2}
 \begin{aligned}
f_{SU(2)}(u) &=\frac{F (u)}{4}\left(\langle \mathfrak{d}|\phi^1\phi^2\rangle+\langle \mathfrak{d}|\phi^2\phi^1\rangle\right)^2\\
&=x_s^2\frac{ u(u-\frac{i}{2})}{(u-i(s-\frac{1}{2}))(u+i(s-\frac{1}{2}))}\frac{x^{+}}{x^{-}}\frac{\sigma_B(u)}{\sigma (u,\bar{u})}\period
 \end{aligned}
 \eeq
Expanding it at weak coupling, we obtain the following expression for $\sqrt{f_{SU(2)}(u)f_{SU(2)}(\bar{u})}$:
\beq
\begin{aligned}
&\sqrt{f_{SU(2)}(u)f_{SU(2)}(\bar{u})}=\\
&\qquad \qquad\frac{x_s^{2}\,  \sqrt{u^2(u^2+\frac{1}{4}})}{4\left(u^2+\left(s-\frac{1}{2}\right)^2\right)}\left[1+\frac{4 g^2}{u^{2}+\frac{1}{4}}\left(\Psi (1+s)+\gamma_E-\log 2\right)+\mathcal{O}(g^4)\right]\period
\end{aligned}
\eeq
 \paragraph{Asymptotic overlap}
 With the two-particle overlap at hand, we can write down the asymptotic overlap formula which includes all the perturbative $1/J$ corrections. The asymptotic formula is expected to be exact up to three loops since the correction to the asymptotic formula (called the wrapping corrections) is known to appear only at four loops. 
 
 To rigorously derive the asymptotic formula, one has to go through steps 3 and 4 in Section \ref{subsec:generalstrategy}: Namely we first formulate the TBA, compute the excited-state $g$-function and finally take the asymptotic limit $J\to \infty$. Such an analysis was done for the structure constants of determinant operators in \cite{Jiang:2019xdz,Jiang:2019zig}. It was then found that the result is given purely in terms of the two-particle overlaps and the Gaudin determinants, and has the same universal structure as the results at weak coupling. Here we {\it assume} that this is the case also for the defect one-point function. Then the asymptotic overlap formula in the $SU(2)$ sector is given by
 \beq\label{eq:asympt}
\left. \frac{\langle \mathcal{D}|{\bf u}\rangle}{\sqrt{\langle {\bf u}|{\bf u}\rangle}}\right|_{\rm asym}=(x_s)^{J}\sqrt{\left(\prod_{m=1}^{M/2}f_{\rm SU(2)}(u_m)f_{\rm SU(2)}(\bar{u}_m)\right)\frac{\det G_{+}}{\det G_{-}}}\comma
 \eeq
 where $|{\bf u}\rangle$ is a parity-symmetric Bethe eigenstate with rapidities ${\bf u}=\{u_1,\bar{u}_1,\ldots, u_{\frac{M}{2}},\bar{u}_{\frac{M}{2}}\}$, and $\det G_{\pm}$ are the Gaudin(-like) determinants whose definitions can be found in e.g.~\cite{Buhl-Mortensen:2017ind,Jiang:2019xdz,Jiang:2019zig}. The prefactor $x_s^{J}$ comes from the rule for pulling out $Z$ fields \eqref{eq:pullingout} and is a counterpart of $i^{J}$ in (1.13) of \cite{Jiang:2019xdz}. We will later give a more physical interpretation of this factor.
 
\begin{figure}[t]
\centering
\includegraphics[clip, height=2.7cm]{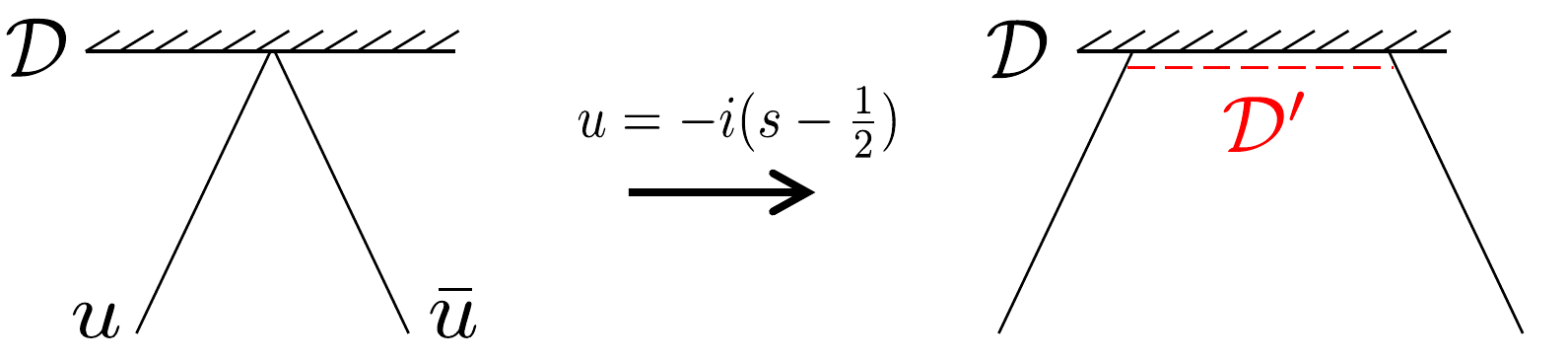}
\caption{Poles of the two-particle overlap and excited boundary states. A pole of the two-particle overlap corresponds to a physical process depicted above, in which a particle gets absorbed by the boundary state and modifies it to an excited boundary state (shown as a red dashed line). Using this property, one can compute the two-particle overlap for the excited boundary state by solving the bootstrap axiom given in Figure \ref{fig:excited}.}\label{fig:poles}
\end{figure}

 Now to perform a comparison with \cite{Buhl-Mortensen:2017ind}, we set $s=(k-1)/2$ and expand \eqref{eq:asympt} at weak coupling. This gives
 \beq\label{eq:oneoverlapfull}
 \left. \frac{\langle \mathcal{D}|{\bf u}\rangle}{\sqrt{\langle {\bf u}|{\bf u}\rangle}}\right|_{\rm asym}=\frac{\left(x_{\frac{(k-1)}{2}}\right)^{L}}{2^{M}}\left[\frac{\sqrt{Q(\frac{i}{2})Q(0)}}{Q(\frac{i (k-2)}{2})}\sqrt{\frac{\det G_{+}}{\det G_{-}}}\mathbb{F}_k+\mathcal{O}(g^{4})\right]\comma
 \eeq
where $Q(u)$ is the Baxter $Q$-function
\beq
Q(u)\equiv\prod_{m=1}^{M}(u-u_m)=\prod_{m=1}^{M/2}(u^2-u_m^2)\comma
\eeq 
and $\mathbb{F}_k$ is the flux factor introduced in \cite{Buhl-Mortensen:2017ind}
\beq
\mathbb{F}_k=1+\left(\Psi \left(\frac{k+1}{2}\right)+\gamma_{E}-\log 2\right)\delta \Delta\comma
\eeq
with $\delta \Delta \equiv \sum_{m=1}^{M}\frac{2g^2}{u_m^{2}+\frac{1}{4}}$. The result \eqref{eq:oneoverlapfull} resembles but does not quite agree with the perturbative answer in \cite{Buhl-Mortensen:2017ind}. The main difference is that, while \eqref{eq:oneoverlapfull} is given by a single term, the result in \cite{Buhl-Mortensen:2017ind} is a sum of $k$ different terms. In what follows, we show that this mismatch can be resolved once we take into account the contributions from excited boundary states.
\paragraph{Excited boundary state} The two-particle overlap $f_{SU(2)}(u)$ given in \eqref{eq:fsu2} has poles at $u=\pm i(s-\frac{1}{2})$. As explained in \cite{Ghoshal:1993tm}, such poles correspond to physical processes in which a particle gets absorbed by the boundary and changes it to an {\it excited boundary state}\footnote{In the boundary scattering picture, this is often called the boundary bound state \cite{Ghoshal:1993tm}. Such boundary bound states also play an important role in the analysis of the spectrum on the so-called $Z=0$ brane in \cite{Hofman:2007xp}.}, see Figure \ref{fig:poles}. This is a boundary scattering version of the relation between poles in the S-matrix and bound states.

\begin{figure}[t]
\centering
\includegraphics[clip,height=2.6cm]{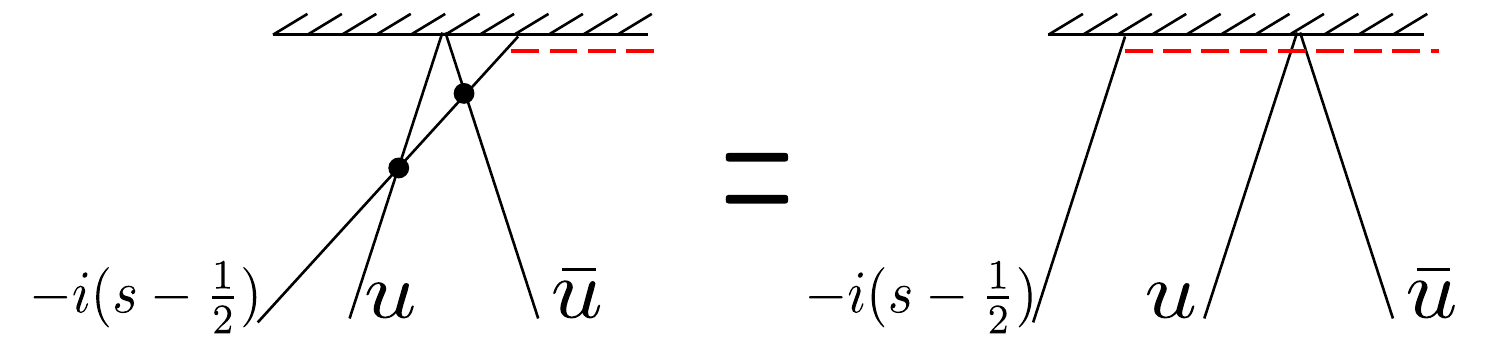}
\caption{The bootstrap axiom for the overlap for the excited boundary state. The overlap for the excited boundary state (depicted on the right hand side) is given by a product of the bulk S-matrices (the black dots in the figure) and the overlap for the original boundary state, as shown on the left hand side of the figure.}\label{fig:excited}
\end{figure}

Excited boundary states can be thought of as bound states of particles and the boundary, and much like usual bound states of particles, we need to include their contributions in order to obtain the correct answer. In integrable field theories, the two-particle overlaps for excited boundary states can be determined from the bootstrap axiom \cite{Ghoshal:1993tm} which is depicted in Figure \ref{fig:excited}. In our case, we can work entirely within the $SU(2)$ sector and the bootstrap axiom gives
\beq\label{eq:howtogetprime}
f^{\prime}_{SU(2)}(u)=\tilde{S}(-i (s-\tfrac{1}{2}),u)\tilde{S}(-i (s-\tfrac{1}{2}),\bar{u})f_{SU(2)}(u)\comma
\eeq
where $f^{\prime}_{SU(2)}(u)\equiv\langle \mathcal{D}^{\prime}|\tilde{\Phi}(u)\tilde{\Phi}(\bar{u})\rangle$ is the two-particle overlap for the excited boundary state $|\mathcal{D}^{\prime}\rangle$, and $\tilde{S}(u,v)$ is the S-matrix in the $SU(2)$ sector (in the string frame) given by \cite{Beisert:2005tm,Beisert:2006qh}
\beq
\tilde{S}(u,v)\equiv\frac{u-v-i}{u-v+i}\frac{x^{+}(u)x^{-}(v)}{x^{-}(u)x^{+}(v)}\frac{1}{(\sigma (u,v))^2}\period
\eeq
We then get\footnote{Here we used the parity invariance of the dressing phase $\sigma (u,v)=\sigma (\bar{v},\bar{u})$.}
\beq\label{eq:excitedf}
\begin{aligned}
f^{\prime}_{SU(2)}(u)=&x_{s-1}^2\frac{x^{+}}{x^{-}}\frac{ u(u-\frac{i}{2})(u-i (s+\frac{1}{2}))(u+i (s+\frac{1}{2}))}{(u-i(s-\frac{1}{2}))(u+i(s-\frac{1}{2}))(u-i (s-\frac{3}{2}))(u+i(s-\frac{3}{2}))}\\
&\times \frac{\sigma_B(u)}{\sigma (u,\bar{u})\left(\sigma (u,i(s-\frac{1}{2}))\sigma (\bar{u},i(s-\frac{1}{2}))\right)^2}\period
\end{aligned}
\eeq
Let us make two remarks before moving on. First the simple relation \eqref{eq:howtogetprime} is valid only in the $SU(2)$ sector, in which the S-matrix and the two-particle overlaps are just scalar factors. In the full $\mathfrak{su}(2|2)^2$ spin chain, one has to consider a matrix analogue of \eqref{eq:howtogetprime}, which is pictorially represented in Figure \ref{fig:excited}. As a result, the two-particle overlap for the excited boundary state will have a different matrix structure from the one for the original boundary state \eqref{eq:matrixstructurestringframe}. Second there is another important difference between the original boundary state and the excited boundary state. In the asymptotic overlap formula  for $|\mathcal{D}\rangle$ \eqref{eq:asympt}, we had an overall factor $x_s^{J}$. Physically this can be interpreted as a propagation factor $e^{ip_{\rm bdy} J}$ with $e^{ip_{\rm bdy}}\equiv x_s$ being the momentum carried by the boundary state $|\mathcal{D}\rangle$. Since the excited boundary state $|\mathcal{D}^{\prime}\rangle$ is a bound state of $|\mathcal{D}\rangle$ and a particle with $u=-i(s-\frac{1}{2})$, the momentum of $|\mathcal{D}^{\prime}\rangle$ is given by
\beq
e^{ip_{\rm bdy}^{\prime}}=e^{ip_{\rm bdy}}\frac{x^{+}(-i(s-\tfrac{1}{2}))}{x^{-}(-i(s-\tfrac{1}{2}))}=x_{s-1}\period
\eeq
This is also consistent with \eqref{eq:excitedf} in which the overall factor $x_s^2$ is replaced by $x_{s-1}^2$.

The excited-state overlap \eqref{eq:excitedf} have new poles at $u=\pm i(s-\frac{3}{2})$. Therefore, by using a particle with rapidity $-i(s-\frac{3}{2})$, we can further excite the boundary state. Setting $s=\frac{ (k-1)}{2}$ and repeating this process until we do not get any new poles, we obtain $k$ different excited boundary states, which we denote by 
\beq
|\mathcal{D}_{k}^{(a)}\rangle\comma\qquad \quad (a=-\tfrac{k-1}{2},\ldots, \tfrac{k-1}{2})\period
\eeq
The two-particle overlaps of these states are given by
\begin{shaded}
\beq
\begin{aligned}
&\tilde{f}_{k}^{(a)}(u)\equiv\langle \mathcal{D}_k^{(a)}|\tilde{\Phi}(u)\tilde{\Phi}(\bar{u})\rangle=x_a^2\frac{x^{+}}{x^{-}}\frac{ u(u-\frac{i}{2})(u^2+\frac{k^2}{4})}{(u^2+\frac{(a-1)^2}{4})(u^2+\frac{(a+1)^2}{4})}\frac{\sigma_B(u)\left(\tilde{\sigma}_{k}^{(a)}(u)\right)^2}{\sigma (u,\bar{u})}\comma
\end{aligned}
\eeq   
with
\beq
\tilde{\sigma}_{k}^{(a)}(u)\equiv\prod_{r=a}^{\frac{k-3}{2}}\frac{1}{\sigma (u, i r)\sigma (\bar{u},ir)}\,.
\eeq
\end{shaded}
\noindent
We can also express $\tilde{\sigma}_{k}^{(a)}$ more explicitly as\footnote{Recall that $x_s$ is given by $x(is)$.}
\beq
\begin{aligned}
\frac{\log \tilde{\sigma}_{k}^{(a)}(u)}{i}=&\chi (x^{-},x_{a})-\chi (x^{+},x_{a})-\chi (x^{-},x_{\frac{ (k-1)}{2}})+\chi (x^{+},x_{\frac{ (k-1)}{2}})\\
&+\chi (-x^{+},x_{a})-\chi (-x^{-},x_{a})-\chi (-x^{+},x_{\frac{ (k-1)}{2}})+\chi (-x^{-},x_{\frac{ (k-1)}{2}})\comma
\end{aligned}
\eeq
where $\chi$ is given by (see e.g.~\cite{Dorey:2007xn,Vieira:2010kb})
\beq
\chi (x,y)\equiv\frac{1}{i}\oint_{|z|=1}\frac{dz}{2\pi i}\oint_{|w|=1}\frac{dw}{2\pi i}\frac{1}{x-z}\frac{1}{y-w}\log \frac{\Gamma [1+i g (z+\frac{1}{z}-w-\frac{1}{w})]}{\Gamma [1-i g (z+\frac{1}{z}-w-\frac{1}{w})]}\period
\eeq

The asymptotic overlap for $|\mathcal{D}_{k}^{(a)}\rangle$ is simply given by replacing $x_s$ and $f_{SU(2)}$ in \eqref{eq:asympt} with $x_{a}$ and $\tilde{f}_{k}^{(a)}$
\beq
\left. \frac{\langle \mathcal{D}_{k}^{(a)}|{\bf u}\rangle}{\sqrt{\langle {\bf u}|{\bf u}\rangle}}\right|_{\rm asym}=(x_{a})^{J}\sqrt{\left(\prod_{m=1}^{M/2}\tilde{f}_{k}^{(a)}(u_m)\tilde{f}_{k}^{(a)}(\bar{u}_m)\right)\frac{\det G_{+}}{\det G_{-}}}\,.
 \eeq

\paragraph{Full asymptotic result} We now propose that the defect one-point function in the asymptotic limit is given by a sum of overlaps for the excited boundary states, namely
\begin{shaded}
\beq
\begin{aligned}\label{eq:fullintegrability1}
\langle \mathcal{O}_{\bf u}(x)\rangle_{\mathcal{D}_k}=\frac{c^{(k)}_{{\bf u}}}{2^{J}\sqrt{L}x_{\perp}^{\Delta}}\comma
\end{aligned}
\eeq
with
\begin{align}\label{eq:fullintegrability2}
c^{(k)}_{{\bf u}}&\equiv\sum_{a=-\frac{k-1}{2}}^{\frac{k-1}{2}}\frac{\langle \mathcal{D}_{k}^{(a)}|{\bf u}\rangle}{\sqrt{\langle {\bf u}|{\bf u}\rangle}}=\mathbb{T}_{k}\frac{\sqrt{Q(\frac{i}{2})Q(0)}}{Q(\frac{i k}{2})}\sqrt{\left(\prod_{m=1}^{M}\sigma_B(u_m)\right)\frac{\det G_{+}}{\det G_{-}}}\comma\\
\mathbb{T}_{k}&\equiv\sum_{a=-\frac{k-1}{2}}^{\frac{k-1}{2}}(x_{a})^{J+M}\frac{\left(Q(\frac{ik}{2})\right)^2}{Q^{-}(i a)Q^{+}(i a)}\prod_{m=1}^{M}\tilde{\sigma}_k^{(a)}(u_m)\period\label{eq:fullintegrability3}
\end{align}
\end{shaded}

Let us make several comments on the formula. Firstly the overall factor $1/2^{J}$ in \eqref{eq:fullintegrability1} is the kinematical factor associated to the $R$-symmetry polarization. For the comparison with the weak-coupling results in the literature, it is often convenient to combine it with the factor $1/2^{M}$ in the boundary dressing phase $\sqrt{\prod_{m}\sigma_B (u_m)}$, and rewrite them as\footnote{Note that $J$ counts the number of $Z$ while $M$ counts the number of $\tilde{\Phi}$. Together they give the length of the operator $L=J+M$.} $1/2^{L}$. Secondly $1/\sqrt{L}$ in \eqref{eq:fullintegrability1} is the usual factor coming from the cyclicity of the trace. See for instance \cite{Escobedo:2010xs} for further explanation. Thirdly $\mathbb{T}_k$ in \eqref{eq:fullintegrability2} is a finite-coupling generalization of the transfer matrix of the Heisenberg spin chain found in the weak-coupling results \cite{Buhl-Mortensen:2015gfd,Buhl-Mortensen:2017ind}. The main difference from  \cite{Buhl-Mortensen:2015gfd,Buhl-Mortensen:2017ind} is the factor $\prod_{m}\tilde{\sigma}_k^{(a)}(u_m)$ which starts to contribute at three loops.  

Our proposal is in perfect agreement\footnote{Precisely speaking the weak-coupling results contain an extra sign factor $(-1)^{L/2}$. (Note that $L$ is even in order for the one-point function to be nonzero.) Keeping track of such an overall sign is practically difficult since the formula contains various square roots, but it would be interesting to clarify this point.} with the results at tree level and one loop in the literature \cite{deLeeuw:2015hxa,Buhl-Mortensen:2015gfd,Buhl-Mortensen:2017ind}, providing strong evidence for the validity of our bootstrap analysis. It will be interesting to perform the higher-loop computation in perturbation theory and compare the results with our predictions.

\section{Discussion and Conclusion}
\label{sec:discussion}

In this paper, we studied the half-BPS superconformal boundary and interface defects of the D5-brane type  in the $\cN=4$ SYM with $U(N)$ gauge group. Defined by unconventional singular Nahm pole configurations of the SYM fields, such defects have only been explored to leading orders in perturbation theory. We presented non-perturbative approaches to this defect CFT problem  based on supersymmetric localization and integrability methods. 

Following the localization setup in \cite{Wang:2020seq}, we have identified the effective 2d defect-Yang-Mills (dYM) theory that captures general ${1\over 16}$-BPS defect observables in the $\cN=4$ SYM with the D5-brane type boundary or interface defect. In particular the dYM contains coupling to a 1d topological quantum mechanics (TQM) which we obtained from localizing the 3d mirror quiver gauge theory description of the Nahm pole boundary condition from \cite{Gaiotto:2008ak}. Insertions of half-BPS operators $\cO_J$ in the SYM with the D5-brane defect translate to insertions of $\tr(\star \cF)^J$ in the dYM where $\cF$ denotes the field strength of the emergent 2d gauge field.  From standard two-dimensional gauge theory techniques,  the latter reduces to a computation in a single matrix model with a novel matrix potential due to the D5-brane defect. Solving this matrix model in the large $N$ limit, we obtained exact defect one-point functions $\la \cO_J \ra_\cD$ in the 't Hooft coupling $\lambda$. Our non-perturbative answers agree with perturbative Feynman diagram computations at weak coupling, and provides a precision test of AdS/CFT with interface defect by comparing with the string theory results on $AdS_5\times S^5$ in the strong coupling regime. 
Going beyond defect observables protected by supersymmetry, we developed a non-perturbative bootstrap-type approach to determining the integrable boundary states in the planar $\cN=4$ SYM corresponding to interface defects, based on recently developed integrability methods of \cite{Jiang:2019xdz,Jiang:2019zig}. The one-point function of a single-trace, generally non-BPS, local operator is then given by the overlap between the integrable boundary state and a closed string state corresponding to the local operator. We explicitly solved the consistency conditions  that define the boundary state, namely the Watson's equation, boundary Yang-Baxter equation and the crossing equations. By dressing the minimal solution with appropriate CDD factors, we obtained defect one-point functions of non-BPS operators (in the $SU(2)$ sector) in the asymptotic limit (large R-charge $J$) that are in perfect agreement with previous results from the traditional integrable spin-chain methods at one-loop. 

There are a number of  future directions worth exploring which we now discuss. First of all, the 2d dYM we have uncovered for the D5-brane defect captures more general observables of the SYM preserving a common supercharge beyond the local operators $\cO_J$ considered in the main text (see  \cite{Wang:2020seq} for the general classification). For example they include correlation functions that involve genuine defect local operators in the TQM sector, as well as ${1\over 8}$-BPS Wilson loops insertions \cite{Drukker:2007yx,Drukker:2007qr} that correspond to ordinary Wilson loops in the dYM.

It will be very interesting  to complete the integrability program we have outlined in Section~\ref{subsec:generalstrategy} by writing down the thermodynamic Bethe ansatz in the open string channel. 
This will enable us to reproduce the BPS defect one-point functions computed in this paper from the supersymmetric localization, and fully determine the non-BPS defect one-point functions at finite 't Hooft coupling in the planar limit.
Combined with the OPE of local operators in the bulk, this gives a way to extract general non-BPS correlation functions in the defect CFT.  Furthermore, boundary (interface) defect crossing symmetry relates factorization channels of correlation functions that exchange  bulk and defect local operators respectively \cite{Liendo:2012hy,Liendo:2016ymz,Billo:2016cpy,deLeeuw:2017dkd}. Therefore, by pursuing a defect superconformal bootstrap program, one can try to determine the full spectrum of operators confined to the defect worldvolume.

Another related direction is to classify the integrable boundary states in accordance with the general construction of interface defects from a chain of $(p,q)$ 5-branes in IIB string theory \cite{Gaiotto:2008sa}. In this paper, we have focused on the case with a single D5-brane and it is more desirable to have a general dictionary\footnote{For a certain class of operators living on the interface which can be described by closed spin chains, a similar question was addressed in \cite{Rapcak:2015lhn}. It studied the action of the dilatation operator perturbatively at one loop, and concluded that the integrability is generally broken for such operators.} between such 5-brane configurations and integrable boundary states. Perhaps a hint towards such a dictionary, these general 5-brane interfaces can be constructed by gluing together  a number of the D5-brane interfaces (at various values of $k$) which we have studied, together with general $SL(2,\mZ)$ duality transformations. This gluing picture is succinctly represented by the matrix model expressions for the defect partition functions using supersymmetric localization (see Section~\ref{sec:simmm}). It would be interesting to understand the corresponding (de)construction of these boundary states on the integrability side.

\section*{Acknowledgement}
We thank Charlotte Kristjansen and Matthias Wilhelm for discussions.
The work of S.K.~is supported by DOE grant number DE-SC0009988.
The work of Y.W. is supported in part by the US
NSF under Grant No. PHY-1620059 and by the Simons Foundation Grant No. 488653.

\bibliographystyle{JHEP}
\bibliography{DefectRef,defREF,SYMdefect}

\end{document}